\documentclass{aa}  

\usepackage{graphicx}
\usepackage{txfonts}

\usepackage{amsmath,bm}
\usepackage{multicol}
\usepackage{subcaption}
\usepackage[colorlinks=true,linkcolor=blue,citecolor=blue]{hyperref}
\usepackage{nccmath}
\usepackage{float}
\usepackage{comment}
\usepackage{mathtools}

\newcommand{\PrM}{{\rm Pr}_{\rm M}}
\newcommand{\ReM}{{\rm Re}_{\rm M}}
\newcommand{\Prot}{P_{\rm rot}}
\newcommand{\Pcyc}{P_{\rm cyc}}
\newcommand{\Ro}{{\rm Ro}}
\newcommand{\Co}{{\rm Co}}
\newcommand{\urms}{{u}_{\rm rms}}

\begin{document} 

\authorrunning{Ortiz-Rodríguez et al.}
\titlerunning{Simulations of dynamo action in slowly rotating M dwarfs}

   \title{Simulations of dynamo action in slowly rotating M dwarfs:\\ Dependence on dimensionless parameters}

   \author{C. A. Ortiz-Rodríguez\inst{1} \and P. J. K\"apyl\"a\inst{2,3,4} \and F. H. Navarrete\inst{5,4}, D. R. G Schleicher\inst{1} \and R. E. Mennickent\inst{1} \and J.P. Hidalgo\inst{1} \and B. Toro \inst{1}}

   \institute{Departamento de Astronom\'ia, Facultad de Ciencias F\'isicas y Matemáticas, Universidad de Concepci\'on, Av. Esteban Iturra s/n Barrio Universitario, Casilla 160-C, Chile \and
%PJK: added
Leibniz-Institut f\"ur Sonnenphysik (KIS), Sch\"oneckstr. 6, 79104 Freiburg, Germany \and
Institut f\"ur Astrophysik und Geophysik, Georg-August-Universit\"at G\"ottingen,
Friedrich-Hund-Platz 1, 37077 G\"ottingen, Germany \and
Nordita, KTH Royal Institute of Technology and Stockholm University, 
10691 Stockholm, Sweden \and
Hamburger Sternwarte, Universit\"at Hamburg, Gojenbergsweg 112, 21029
Hamburg, Germany}

   \date{}

% \abstract{}{}{}{}{} 
% 5 {} token are mandatory
 
  \abstract
   {}
   {The aim of this study is to explore the magnetic and flow
     properties of fully convective M dwarfs as a function of
rotation period $\Prot$ and magnetic {Reynolds $\ReM$ and} Prandlt number{s} $\PrM$.}
   {We performed three-dimensional simulations of fully convective
     stars using a star-in-a-box setup. This setup allows global
     dynamo
     simulations in a sphere embedded in a Cartesian cube.
     The equations of non-ideal magnetohydrodynamics were
     solved with the {\sc Pencil Code}. We used the stellar parameters of
     an M5 dwarf with 0.21$M_{\odot}$ at three rotation {rates
       corresponding to rotation} periods ($\Prot$): 43,
     61 and 90 days, and varied the magnetic Prandtl number in the range
     from 0.1 to 10.}
   {{We found systematic differences in the behaviour of the
       large-scale magnetic field as functions of rotation and
       $\PrM$}. For the simulations with $\Prot=43$
     days and $\PrM \leq 2$, we found cyclic large-scale magnetic
     fields. For $\PrM > 2$ the cycles vanish and field shows
     irregular reversals. In simulations with $\Prot=61$
     days for $\PrM \leq 2$ the cycles are less clear and the
     reversal are less periodic. In the higher-$\PrM$ cases, the
     axisymmetric mean field {shows irregular variations}. For the
     slowest
     rotation case with $\Prot = 90$~days, the field {has an important dipolar component}
%PJK: Is this clear from the power spectra, i.e. that the l=1 mode is
%PJK: dominant?
     for $\PrM \leq
%PJK
     5$. For the highest $\PrM$ the large-scale magnetic field is  
     predominantly irregular at mid-latitudes, with
     quasi-stationary fields near the poles. {For the simulations with cycles, the cycle period length slightly increases with increasing $\ReM$.}}
%PJK: We should also mention something about magnetic field saturation
%PJK: levels and cycle periods.
   {}

%   \keywords{
%               }


   \maketitle
%
%-------------------------------------------------------------------

\section{Introduction}

Magnetic fields in stars have been studied both theoretically and
through observations, particularly magnetic fields of {solar-type
main-sequence stars \citep[e.g.][and references
  therein]{2017LRSP...14....4B}}. M dwarfs are low-mass
main-sequence stars
with a structure that undergoes a transition from fully convective for
masses up to $0.35M_{\odot}$ to a solar-like structure (radiative core
and convective envelope) for {higher mass}
\citep{1997A&A...327.1039C}. These stars are found to be
magnetically active, as shown by \citet{Saar1985} where surface
magnetic activity was confirmed for M dwarfs with infrared
measurements. Today there is considerable observational evidence of
magnetic activity in M dwarfs that show magnetic field strengths
reaching up to a few kG \citep[see][and references
  therein]{kochukhov2021magnetic}. Because of the lack of a
tachocline, {the shear layer between the radiative and convective zones,} fully convective M dwarfs are quite interesting from the
point of view of dynamo theory and can help us to understand whether
a tachocline has a strong impact on the dynamo
itself. In this context, \citet{wright2016solar} reported that the
X-ray emission of {fully and partially} convective stars follows a
similar trend with the Rossby number {$\Ro = \Prot/\tau$, which is
  the ratio of the rotation period and convective turnover time, and
  which measures the rotational influence} on convective flows. It was
found that
the X-ray emission increases with decreasing $\Ro$ until
$\Ro\approx0.1$, and for smaller Ro the X-ray luminosity saturates.
Furthermore, \citet{newton2017halpha} found a similar trend, a
saturated relation between the chromospheric H$\alpha$ emission and
$\Ro$ for rapidly rotating M dwarfs and a power-law decay of the
H$\alpha$ emission with increasing $\Ro$ for slowly rotating
stars. The transition occurs near $\Ro = 0.2$. In addition, {Doppler and Zeeman-Doppler inversions have revealed that fully
  convective M dwarfs often show large-scale magnetic fields and that
  for rapid enough rotation both dipolar and multipolar fields are
  possible
  \citep[e.g.][]{2010MNRAS.407.2269M,kochukhov2021magnetic}. Furthermore,}
\citet{klein2021large} {found} that the fully convective star Proxima
Centauri has a seven year activity cycle.

Numerical simulations of stars are performed to achieve a better
understanding of their magnetic fields, dynamos, and convection as
functions of stellar parameters and dimensionless {quantities},
such as the
magnetic Prandtl number, which is an intrinsic property of the fluid
defined by the ratio of {kinematic} viscosity $\nu$ and
resistivity $\eta$ of the
plasma. Some authors have performed magnetohydrodynamic (MHD)
simulations of fully convective M dwarfs, which are particularly
interesting for comparison with solar dynamo models due to the lack of
a tachocline.
{The first simulations of fully convective M dwarfs were presented by
\cite{dobler2006}, who used a star-in-a-box model to study dynamos as
a function of rotation. They found predominantly quasi-static
large-scale magnetic fields and typically weak or anti-solar
differential rotation with faster poles and slower equator. These
simulations had relatively modest fluid and magnetic Reynolds numbers
as well as low density stratification.}
%PJK
\citet{browning2008simulations} presented simulations of fully
convective M dwarfs using the anelastic magnetohydrodynamic equations,
considering a spherical domain extending from 0.08 to 0.96 stellar
radius, finding magnetic fields with significant axisymmetric
components. In simulations without magnetic fields, the differential
rotation is strong and solar-like {with fast equator and slow
  poles}, while in magnetic simulations it is
reduced, and tends to a solid body rotation in the most turbulent
magnetohydrodynamical simulations.
{A similar numerical approach was taken in the studies of
  \cite{YCMGRPW15} and \cite{2016ApJ...833L..28Y} who used strongly
  stratified anelastic simulations to study the coexistence of dipolar
  and multipolar dynamos and cyclic solutions at relatively slow
  rotation corresponding to parameter regime similar to Proxima
  Centauri, respectively.}
{More recently,} \citet{brown2020single} performed simulations of
fully convective M dwarfs in spherical coordinates, finding {cyclic hemispheric} dynamos in their models.

The rotation period of the star, $\Prot$, is a key factor that
determines the nature of the dynamo. This is evidenced by
observational studies of M dwarfs, which demonstrate that with
decreasing $\Prot$ {the magnetic field strength increases}
\citep[e.g.][]{wright2018stellar,2022A&A...662A..41R}. This has also
been shown numerically by, {for example}, \citet{kaepylae2021}
{who used} a star-in-a-box
model for fully convective stars {and found} increasing magnetic
field
strength with decreasing rotation period. Furthermore, different
dynamo {modes were found as a function of rotation} in that
work. Slowly rotating stars have mostly {axisymmetric and
  quasi-steady large-scale magnetic fields}, for intermediate rotation
the large-scale field is mostly axisymmetric and cyclic, and in the
case of rapid rotation, the large-scale magnetic fields are
predominantly non-axisymmetric {with a dominant $m=1$ mode}. As
demonstrated by
\citet{kaepylae2021}, the large-scale dynamo is sustained even in the
absence of a tachocline. In this sense, the work by
\citet{bice2020probing} using simulations of early M dwarfs supports
the hypothesis that the tachocline is not necessary for producing
strong toroidal magnetic fields, although it may {generate} stronger
fields in faster rotators.

In this paper, we present three-dimensional MHD simulations of fully
convective M dwarfs with the star-in-a-box setup described in
\citet{kaepylae2021} \citep[see also][]{dobler2006}. Our main goal is
to explore the dependence on dimensionless parameters, in particular
the magnetic Prandtl {and Reynolds numbers $\PrM$ and $\ReM$, which are crucial ingredients}
for dynamos and plasmas in general. High and low values of
$\PrM$ {and $\ReM$} lead to very different dynamo scenarios; at
low $\PrM$ the
magnetic energy is {dissipated in the inertial range of the flow
  and small-scale dynamo action requires a much higher $\ReM$ to be
  excited \citep[e.g.][]{SICMPY07,2018AN....339..127K}. On the other
  hand, stars typically have $\PrM\ll1$ and $\ReM\gg1$
  \citep[e.g.][]{2019ApJ...876...83A,2022ApJS..262...19J}.} Our
simulations were performed for a set of
rotation periods $\Prot$ ranging from 43 to 90 days, the latter 
being the rotation period of Proxima Centauri, and for values of $\PrM$ {and $\ReM$}
ranging from 0.1 to 10 {and 21 to over 1400, respectively,} which
is the numerically feasible range for this type
of simulations. The methods and model are described in Section \ref{methods},
while the description and analysis of the results is provided in Section
\ref{results}. We discuss the conclusions in Section \ref{conclusions}.
%--------------------------------------------------------------------
\section{Methods}\label{methods}
\subsection{The model}\label{model}

We use the star-in-a-box model described in \citet{kaepylae2021},
which is based on the set-up of \citet{dobler2006}. The model allows
dynamo simulations of {entire stars}. In the present scenario,
we use a sphere of radius $R$ that is enclosed in a cube with side
$2.2\,R$. We solve the induction, continuity, momentum,
and energy conservation equations:
\begin{ceqn}
    \begin{eqnarray}
        \frac{\partial \bm{A}}{\partial t} & = & \bm{u} \times \bm{B} - \eta \mu_0 \bm{J},
        \\
        \frac{{\rm D} \ln \rho}{{\rm D} t} & = & -\bm{\nabla} \bm\cdot \bm{u},
        \\
        \frac{{\rm D} \bm{u}}{{\rm D} t}  &= & -\bm{\nabla} \Phi - \frac{1}{\rho}        \left(\bm{\nabla}p - \bm{\nabla}\!\bm\cdot\!2 \nu \rho\bm{\mathsf{S}} \!+\!\bm{J}
        \times \bm{B} \right)\!-\!2 \bm{\Omega} \times \bm{u}\!+\!\bm{f}_d,
        \\
%        T \frac{{\rm D} s}{{\rm D} t} & = & - \frac{1}{\rho}\left[\bm{\nabla}\bm\cdot
%        \left(\bm{F}_{\rm rad}\!+\!\bm{F}_{\rm SGS} \right) + \mathcal{H - C}
%PJK: resolved sign error for C and H
        T \frac{{\rm D} s}{{\rm D} t} & = & \frac{1}{\rho}\left[\mathcal{H - C} - \bm{\nabla}\bm\cdot
        \left(\bm{F}_{\rm rad}\!+\!\bm{F}_{\rm SGS} \right)
%        \right] + 2\nu\bm{\mathsf{S}}^2 + \mu_0 \eta \bm{J}^2,
%FHN: 1/rho is missing
        \right] + 2\nu\bm{\mathsf{S}}^2 + \frac{\mu_0 \eta \bm{J}^2}{\rho},
    \end{eqnarray}
\end{ceqn}
where $\bm{A}$ is the magnetic vector potential, $\bm{u}$ is the velocity
field, $\bm{B} = \bm{\nabla} \times \bm{A}$ is the magnetic field,
$\mu_0$ is the magnetic permeability of vacuum, $\eta$ is the magnetic 
diffusivity, $\rho$ is the density of the fluid, ${\rm D}/{\rm D}t =
\partial/\partial t + {\bm u} \bm\cdot \bm{\nabla}$ is the advective
derivative, $T$ is the temperature, $\Phi$ is the gravitational
potential, $p$ is the pressure, $\nu$ is the kinematic viscosity, $s$
is the specific entropy, $\bm{J} = \bm{\nabla} \times \bm{B}/\mu_0$ is
the current density, $\bm{\Omega} = \Omega_0 \hat{\bm z}$ is the
rotation vector, with $\Omega_0$ being the {mean} angular velocity
of the star and
$\hat{\bm z}$ the unit vector along the rotation axis, and
$\bm{\mathsf{S}}$ is the traceless rate-of-strain tensor,
\begin{ceqn}
    \begin{equation}
%        \mathsf{S}_{ij} = \frac{1}{2}(u_{i,j} + u_{j,i}) - \frac{1}{3} \delta_{ij} \boldsymbol{\nabla} \cdot \boldsymbol{u}
%PJK: \bm, comma
      \mathsf{S}_{ij} = \frac{1}{2}(u_{i,j} + u_{j,i}) - \frac{1}{3} \delta_{ij} \boldsymbol{\nabla} \bm\cdot \boldsymbol{u},
    \end{equation}
\end{ceqn}
where the commas denote differentiation. $\mathcal{H}$ and
$\mathcal{C}$ describe heating and cooling, and $\boldsymbol{f}_d$
describes the damping of flows outside the star {\citep[see][for more
  details]{kaepylae2021}}. The radiative flux is
%PJK
given by
\begin{ceqn}
\begin{equation}
	\bm{F}_{\rm rad}  =  - K \bm{\nabla}T,
\end{equation}
\end{ceqn}
where $K$ corresponds to Kramers opacity law, {where its
  powerlaw exponents are the same of} \cite{kaepylae2021}. The
subgrid-scale (SGS) entropy flux $\bm{F}_{\rm SGS}$ damps fluctuations
near the grid
scale, but contributes {only negligibly} to the net energy
transport. It is
given by
\begin{ceqn}
\begin{equation}
    \bm{F}_{\rm SGS}  =  - \chi_{\rm SGS} \rho \bm{\nabla} s',
\end{equation}
\end{ceqn}
where $\chi_{\rm SGS}$ is the SGS diffusion coefficient,
$s' = s - \bar{s}_t$ is the entropy fluctuation and $\bar{s}_t$ is a 
running temporal mean of the entropy.
{Note that the SGS flux used here does not include the temperature
  $T$. This form of the SGS flux is appropriate if the entropy
  equation is solved, whereas the $T$ factor appears in the SGS term
  if the corresponding energy equation was solved
  \citep{2015JPlPh..81e3904R}.}

The simulations were run with the {\sc Pencil
  Code}\footnote{\href{https://github.com/pencil-code}{https://github.com/pencil-code}}
\citep{2021JOSS....6.2807P}, which is a high-order finite-difference
code for solving partial differential equations with primary
applications in compressible astrophysical magnetohydrodynamics (MHD).

\subsubsection{Dimensionless parameters}
Each simulation is characterized by various dimensionless numbers. 
These parameters are usually order of magnitude ratios of {various} terms in the MHD equations {or of the corresponding
  timescales}.

{The effect of rotation relative to viscosity is measured by the}
Taylor number, given by
\begin{ceqn}
\begin{equation}
    {\rm Ta} = \frac{4 \Omega_0^2 R^4}{\nu^2}.
\end{equation}
The Coriolis number is a measure of the influence of rotation on the
flow
\begin{equation}\label{eq:coriolis}
    {\rm Co} = \frac{2 \Omega_0}{u_{\rm rms}k_R},
\end{equation}
where $u_{\rm rms}$ is the volume-averaged root-mean-square velocity and
$k_R = 2 \pi / R$ is the scale of the largest convective
eddies. {Another definition of the Coriolis number used in
  other studies \citep[e.g.][]{brown2020single, kaepylae2021} is based
  on the vorticity, and considers the local lengthscale. This is
  defined by}
\begin{equation}\label{eq:coriolis_omega}
{\rm Co}_{\omega} = \frac{2\Omega_0}{\omega_{\rm rms}},
\end{equation}
{where} ${\omega_{\rm rms}}$ {is the volume averaged rms
  vorticity, with} $\bm\omega = \bm\nabla \times {\bm u}$. The fluid and
magnetic Reynolds numbers, SGS and magnetic Prandtl, and SGS P\'eclet
numbers are defined as
\begin{equation}
    {\rm Re} = \frac{u_{\rm rms}}{\nu k_{R}}, \; \; \; \; {\rm Re}_{\rm M} = \frac{u_{\rm rms}}{\eta k_R},
\end{equation}
\begin{equation}
    {\rm Pr}_{\rm SGS} = \frac{\nu}{\chi_{\rm SGS}}, \; \; \; {\rm Pr}_{\rm M} = \frac{\nu}{\eta}, \; \; \; {\rm Pe} = \frac{u_{\rm rms}}{\chi_{\rm SGS} k_R}.
\end{equation}
\end{ceqn}

\subsubsection{Physical units {and nondimensional quantities}}\label{sub:units}

We model a main-sequence (M5) dwarf with the same stellar parameters
as in \citet{dobler2006} and \citet{kaepylae2021}. The mass, radius,
and luminosity of the star are $M_{\star} = 0.21 M_{\odot}$,
$R_{\star} = 0.27 R_{\odot}$, and $L_{\star} = 0.008 L_{\odot}$,
respectively. We use an enhanced luminosity approach
\citep{kapyla2020sensitivity} to reduce the gap between the thermal
and dynamical timescales, such that fully compressible simulations are
feasible. This implies that the results need to be scaled suitably
for comparison with real stars. The conversion factor between the
rotation rate, {length, time, velocity, and magnetic
  fields} in the simulation and in physical units {are the same
  as those used by \cite{kaepylae2021} (see their Appendix~A). 
  Nondimensional quantities are obtained by using the stellar radius
  as the unit of length $[x] = R$. Time is given in terms of the
  free-fall time $\tau_{\rm ff} = \sqrt{R^3/GM}$, the unit of velocity
  is $[U] = R/\tau_{\rm ff}$, and magnetic
  fields are
  given in terms of the equipartition field strength $B_{\rm eq} =
  \langle \sqrt{\mu_0 \rho \bm{U}^2} \rangle$, where $\langle
  . \rangle$ stands for time and volume averaging. }

%%%%%%%%%%%%%%%%%%%%%%%%%%%%%%%%%%%%%%%%%%
\section{Results}\label{results}

We present a set of 3D MHD simulations in the
slow to intermediate rotation regime {with global Coriolis number  $\Co$
  ranging between 3.1 and 12.9} (see
Table~\ref{tab:summary_table}). {The rotation rates are $\tilde{\Omega} = 1.0$, $0.7$ and $0.5$ (which correspond to
$\Prot = 43$, $61$, and $90$~days) in sets A, B, and C, respectively.}
The magnetic
Prandtl number $\PrM$ varies set between $0.1$ and $10$ {($0.5$
  and $10$) in set A (sets B and C)}.

\begin{table*}[h!]
\caption{Simulation parameters.}
%\resizebox{\textwidth}{!}{%
\centering
\label{tab:summary_table}
\begin{tabular}{ccccccccccccc}
\hline
Sim  & ${{\tilde {\Omega}}}$ & $\Tilde{u}_{\rm rms}$ & $B_{\rm rms}$\,$[B_{\rm eq}]$ & $\PrM$       & ${\rm Pr_{SGS}}$ & ${\rm Re_M}$  & ${\rm Re}$   & ${\rm Co}$    & ${\bf{\rm Co_{\omega}}}$ & ${\rm Ta}$                      & ${\rm Pe}$  & Grid    \\ \hline
A1 & ${1.0}$ & ${0.022}$ & ${0.92}$ & 0.1 & 0.04 & ${55}$ & ${549}$ & ${9.2}$ & ${1.4}$ & ${4.00 \cdot 10^{10}}$ & ${22}$ & $200^3$ \\
A2 & ${1.0}$ & ${0.021}$ & ${0.94}$ & 0.1 & 0.04 & ${79}$ & ${788}$ & ${9.3}$ & ${1.2}$ & ${8.30 \cdot 10^{10}}$ & ${32}$ & $288^3$ \\
A3 & ${1.0}$ & ${0.021}$ & ${0.91}$ & 0.2 & 0.08 & ${54}$ & ${ 272}$ & ${9.3}$ & ${1.6}$ & ${1.00 \cdot 10^{10}}$ & ${22}$ & $200^3$ \\
A4 & ${1.0}$ & ${0.022}$ & ${0.81}$ & 0.5 & 0.20 & ${54}$ & ${109}$ & ${9.3}$ & ${1.9}$ & ${1.60 \cdot 10^9}$ & ${22}$ & $200^3$ \\
A5 & ${1.0}$ & ${0.022}$ & ${0.75}$ & 0.7 & 0.28 & ${55}$ & ${78}$ & ${9.2}$ & ${2.0}$ & ${8.16 \cdot 10^8}$ & ${22}$ & $200^3$ \\
A6 & ${1.0}$ & ${0.020}$ & ${0.91}$ & 0.7 & 0.28 & ${75}$ & ${107}$ & ${9.7}$ & ${2.0}$ & ${1.69 \cdot 10^9}$ & ${30}$ & $200^3$ \\
A7 & ${1.0}$ & ${0.021}$ & ${0.84}$ & 0.9 & 0.20 & ${54}$ & ${108}$ & ${9.4}$ & ${1.9}$ & ${1.60 \cdot 10^9}$ & ${22}$ & $200^3$ \\
A8* & ${1.0}$ & ${0.030}$ & $-$ & 0.2 & 0.28 & ${22}$ & ${108}$ & ${6.6}$ & ${1.8}$ & ${8.16 \cdot 10^{8}}$ & ${30}$ & $200^3$ \\
A9 & ${1.0}$ & ${0.022}$ & ${0.71}$ & 0.5 & 0.28 & ${39}$ & ${78}$ & ${9.3}$ & ${2.1}$ & ${8.16 \cdot 10^{8}}$ & ${22}$ & $200^3$ \\
A10 & ${1.0}$ & ${0.020}$ & ${0.89}$ & ${1.0}$ & 0.20 & ${105}$ & ${105}$ & ${9.7}$ & ${1.9}$ & ${1.60 \cdot 10^9}$ & ${21}$ & $200^3$ \\
A11 & ${1.0}$ & ${0.020}$ & ${0.88}$ & ${1.0}$ & 0.28 & ${73}$ & ${73}$ & ${9.9}$ & ${2.1}$ & ${8.16 \cdot 10^8}$ & ${20}$ & $200^3$ \\
A12* & ${1.0}$ & ${0.028}$ & $-$ & ${1.0}$ & 0.40 & ${70}$ & ${70}$ & ${7.2}$ & ${2.0}$ & ${4.00 \cdot 10^8}$ & ${28}$ & $200^3$ \\
A13 & ${1.0}$ & ${0.019}$ & ${0.81}$ & ${2.0}$ & 0.40 & ${99}$ & ${50}$ & ${10.1}$ & ${2.3}$ & ${4.00 \cdot 10^8}$ & ${20}$ & $200^3$ \\
A14 & ${1.0}$ & ${0.017}$ & ${1.20}$ & ${5.0}$ & 0.40 & ${208}$ & ${42}$ & ${12.1}$ & ${2.4}$ & ${4.00 \cdot 10^8}$ & ${17}$ & $200^3$ \\
A15 & ${1.0}$ & ${0.017}$ & ${1.16}$ & ${7.0}$ & 0.40 & ${300}$ & ${42}$ & ${12.0}$ & ${2.5}$ & ${4.00 \cdot 10^8}$ & ${17}$ & $200^3$ \\
A16 & ${1.0}$ & ${0.016}$ & ${1.24}$ & ${10.0}$ & 0.40 & ${390}$ & ${39}$ & ${12.9}$ & ${2.5}$ & ${4.00 \cdot 10^8}$ & ${16}$ & $200^3$ \\ \hline
B1 & ${0.7}$ & ${0.024}$ & ${0.68}$ & 0.5 & 0.40 & ${84}$ & ${167}$ & ${5.9}$ & ${1.1}$ & ${1.51 \cdot 10^9}$ & ${62}$ & $200^3$ \\
B2 & ${0.7}$ & ${0.024}$ & ${0.74}$ & ${1.0}$ & 0.40 & ${168}$ & ${168}$ & ${5.8}$ & ${1.0}$ & ${1.51 \cdot 10^9}$ & ${68}$ & $576^3$ \\
B3 & ${0.7}$ & ${0.022}$ & ${0.88}$ & ${2}$ & 0.40 & ${315}$ & ${158}$ & ${6.2}$ & ${1.0}$ & ${1.51 \cdot 10^9}$ & ${64}$ & $576^3$ \\
B4 & ${0.7}$ & ${0.020}$ & ${1.03}$ & ${5}$ & 0.40 & ${714}$ & ${143}$ & ${6.9}$ & ${1.1}$ & ${1.51 \cdot 10^9}$ & ${58}$ & $576^3$ \\
B5 & ${0.7}$ & ${0.019}$ & ${1.02}$ & ${10}$ & 0.40 & ${1360}$ & ${135}$ & ${7.2}$ & ${1.2}$ & ${1.51 \cdot 10^9}$ & ${55}$ & $576^3$ \\ \hline
C1 & ${0.5}$ & ${0.025}$ & ${0.80}$ & ${1.0}$ & 0.20 & ${256}$ & ${256}$ & ${3.9}$ & ${0.6}$ & ${1.60\cdot 10^9}$ & ${51}$ & $576^3$ \\
C2* & ${0.5}$ & ${0.032}$ & $-$ & 0.5 & 0.40 & ${41}$ & ${21}$ & ${3.1}$ & ${0.9}$ & ${1.00 \cdot 10^8}$ & ${32}$ & $200^3$ \\
C3 & ${0.5}$ & ${0.026}$ & ${0.53}$ & ${1.0}$ & 0.40 & ${67}$ & ${67}$ & ${3.8}$ & ${1.1}$ & ${1.00 \cdot 10^8}$ & ${27}$ & $200^3$ \\
C4 & ${0.5}$ & ${0.024}$ & ${0.85}$ & ${2}$ & 0.40 & ${337}$ & ${168}$ & ${4.1}$ & ${0.7}$ & ${7.71 \cdot 10^8}$ & ${68}$ & $576^3$ \\
C5 & ${0.5}$ & ${0.021}$ & ${1.17}$ & ${5}$ & 0.40 & ${750}$ & ${150}$ & ${4.7}$ & ${0.7}$ & ${7.71 \cdot 10^8}$ & ${60}$ & $576^3$ \\
C6 & ${0.5}$ & ${0.020}$ & ${1.02}$ & ${10}$ & 0.40 & ${1419}$ & ${142}$ & ${4.9}$ & ${0.8}$ & ${7.71 \cdot 10^8}$ & ${51}$ & $576^3$ \\ \hline
\end{tabular}
\tablefoot{Summary of the simulations. From left to right the columns
  correspond to the simulation name, {$\tilde{\Omega} = \Omega \tau_{\rm ff}$ is the normalized rotation rate, $\tilde{u}_{\rm rms} =
    \urms/(GM/R)^{1/2}$ is the normalized} root-main-square velocity,
  {$B_{\rm rms}$ is the}
  root-main-square magnetic field strength {in units of the
    equipartition strength}, {$\PrM$ and $\Pr_{\rm SGS}$ are the}
  magnetic and
  sub-grid-scale Prandtl numbers, {$\ReM$ and ${\rm Re}$ are the} magnetic and fluid Reynolds numbers,
  {${\rm Co}$ and ${\rm Co}_{\omega}$ are the global and local
    Coriolis numbers}, {${\rm Ta}$ is the Taylor number} and {${\rm Pe}$ is the P\'eclet number.} The last column indicates the grid resolution.
  Asterisks indicate runs with no dynamo.}
\end{table*}

\subsection{Flow properties}

%PJK: Introduced a new subsection
\subsubsection{Differential rotation and meridional circulation}

The averaged rotation rate in cylindrical coordinates is given {by}
\begin{equation}
    \overline{\Omega}(\varpi,z) = \Omega_0 + \overline{U}_{\phi}(\varpi,z)/\varpi,
\end{equation}
where $\varpi = r \sin{\theta}$ is the cylindrical radius, {and
  where the overbar denotes azimuthal averaging}. The averaged
meridional flow is given {by}
\begin{equation}\label{eq:meridional_flow}
    \overline{U}_{\rm mer} (\varpi, z) = (\overline{U}_{\varpi}, 0, \overline{U}_z). 
\end{equation}
The angular velocity does not only vary with depth but also with
latitude. {A} way to quantify this is by measuring the amplitude of
the {radial and latitudinal differential rotation with}
\begin{ceqn}
\begin{equation}\label{eq:radial}
    \Delta_{\Omega}^{(r)} = \frac{\overline{\Omega}_{\rm top, eq} - \overline{\Omega}_{\rm bot, eq}}{\overline{\Omega}_{\rm top, eq}},
\end{equation}
\begin{equation}\label{eq:latitudinal}
    \Delta_{\Omega}^{(\overline{\theta})} = \frac{\overline{\Omega}_{\rm top, eq} - \overline{\Omega}_{\rm top, \overline{\theta}}}{\overline{\Omega}_{\rm top, eq}}
\end{equation}
\end{ceqn}
where the subscripts {top, bot, eq, and $\overline{\theta}$
  correspond to $R = 0.9R$, $r = 0.1R$, $\theta = 0\degr$, and an
  average of $\overline{\Omega}$ for latitudes $+\theta$ and $-\theta$
  in spherical coordinates, respectively}.

\begin{table}[!htbp]
\centering
\begin{tabular}{ccccc}
\hline
Sim & $\Delta_{\Omega}^{(r)}$ & $\Delta_{\Omega}^{(\overline{\theta})}(60^{\circ})$ & $\Delta_{\Omega}^{(\overline{\theta})}(75^{\circ})$ & $\tilde{\overline{U}}_{\rm mer}^{\rm rms}$\\ 
\hline
A1   & 0.13   & 0.038   & 0.044   & $1.8 \cdot 10^{-3}$   \\
A2   & 0.13   & 0.035   & 0.044   & $1.2 \cdot 10^{-3}$   \\
A3   & 0.13   & 0.039   & 0.046   & $1.9 \cdot 10^{-3}$   \\
A4   & 0.15   & 0.044   & 0.054   & $1.9 \cdot 10^{-3}$   \\
A5   & 0.17   & 0.052   & 0.063   & $2.0 \cdot 10^{-3}$   \\
A6   & 0.12   & 0.037   & 0.049   & $1.2 \cdot 10^{-3}$   \\
A7   & 0.14   & 0.041   & 0.052   & $1.9 \cdot 10^{-3}$   \\
A8*  & 0.28   & 0.100   & 0.100   & $2.5 \cdot 10^{-3}$   \\
A9   & 0.18   & 0.060   & 0.069   & $2.0 \cdot 10^{-3}$   \\
A10  & 0.14   & 0.036   & 0.048   & $1.9 \cdot 10^{-3}$   \\
A11  & 0.12   & 0.036   & 0.046   & $1.9 \cdot 10^{-3}$   \\
A12* & 0.23   & 0.092   & 0.091   & $2.4 \cdot 10^{-3}$   \\
A13  & 0.11   & 0.037   & 0.046   & $1.2 \cdot 10^{-3}$   \\
A14  & 0.020  & 0.018   & 0.019   & $1.3 \cdot 10^{-3}$   \\
A15  & 0.015  & 0.023   & 0.025   & $1.3 \cdot 10^{-3}$   \\
A16  & -0.006 & 0.019   & 0.017   & $1.1 \cdot 10^{-3}$   \\ \hline
B1   & 0.067  & -0.006  & -0.020  & $1.4 \cdot 10^{-3}$   \\
B2   & 0.165  & 0.073   & 0.093   & $2.4 \cdot 10^{-3}$   \\
B3   & 0.100  & 0.062   & 0.076   & $1.8 \cdot 10^{-3}$   \\
B4   & 0.036  & 0.051   & 0.057   & $1.6 \cdot 10^{-3}$   \\
B5   & -0.029 & 0.041   & 0.043   & $1.1 \cdot 10^{-3}$   \\ \hline
C1   & 0.122  & 0.104   & 0.122   & $1.8 \cdot 10^{-3}$   \\
C2*   & 0.087  & 0.067   & 0.060   & $6.7 \cdot 10^{-4}$   \\
C3   & 0.008  & -0.043  & -0.069  & $1.4 \cdot 10^{-3}$   \\
C4   & 0.052  & 0.095   & 0.101   & $1.7 \cdot 10^{-3}$   \\
C5   & -0.013 & 0.086   & 0.078   & $1.6 \cdot 10^{-3}$   \\
C6   & -0.058 & 0.064   & 0.063   & $2.0 \cdot 10^{-3}$   \\ \hline
\end{tabular}
\caption{Amplitudes of the temporally and azimuthally averaged angular 
velocity $\overline{\Omega}(r, \theta)$. From left to right: the name
of the simulation, the amplitudes of the
radial and latitudinal differential rotation at $60^{\circ}$ and
$75^{\circ}$ {according to in Eqs.~(\ref{eq:radial}) and
  (\ref{eq:latitudinal})}, respectively, and the rms value of the
meridional flow speed $\tilde{\overline{U}}_{\rm mer}^{\rm rms} =
(GM/R)^{-1/2}(\overline{U}_{\varpi}^2 + \overline{U}_z^2)^{1/2}$.}
\label{tab:amplitude}
\end{table}

\begin{figure}[t!]
\includegraphics[width=\linewidth]{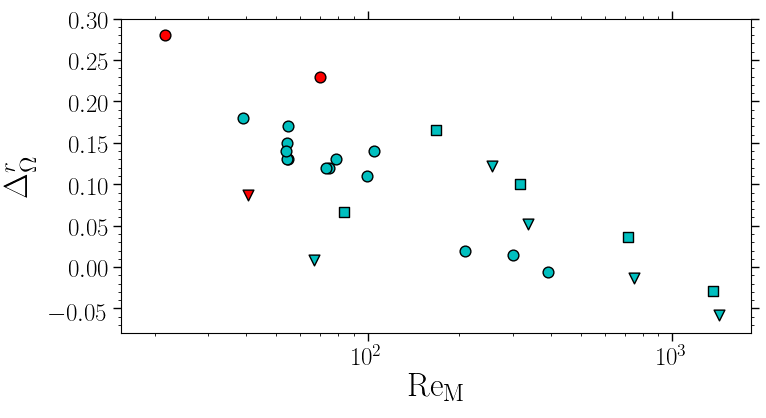}
\caption{Amplitude of the radial differential rotation as a function
  of the magnetic Reynolds number for simulations of set A (circles), B (squares) and C (triangles). Cyan (red) is for simulations with dynamo (without
  dynamo).}
\label{fig:deltaomr_rem}
\end{figure}

\begin{figure*}[h!]
\centering
\includegraphics[width=0.33\textwidth]{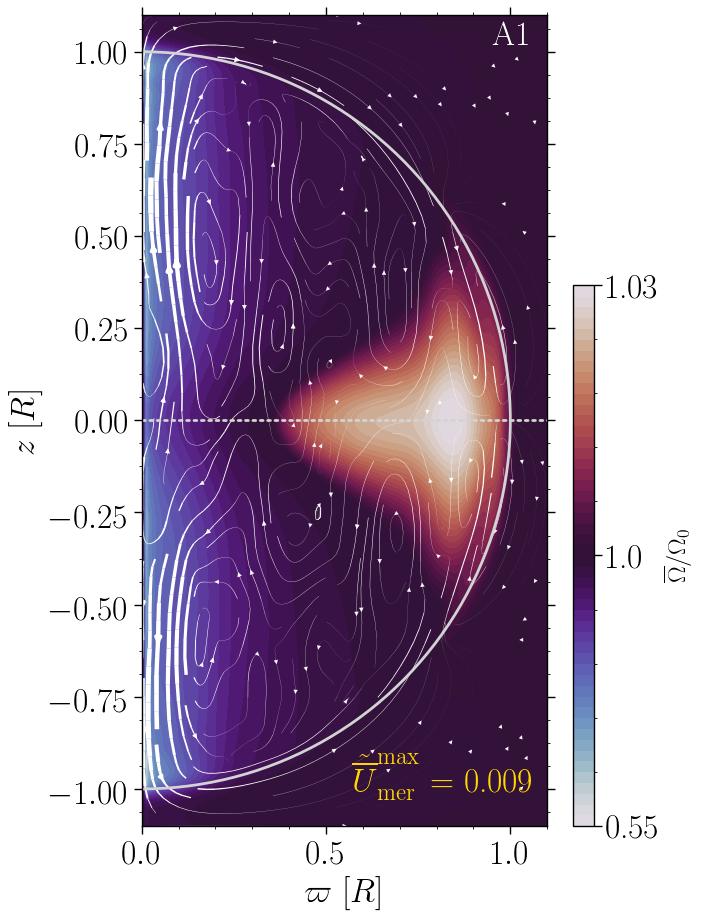}
\includegraphics[width=0.33\textwidth]{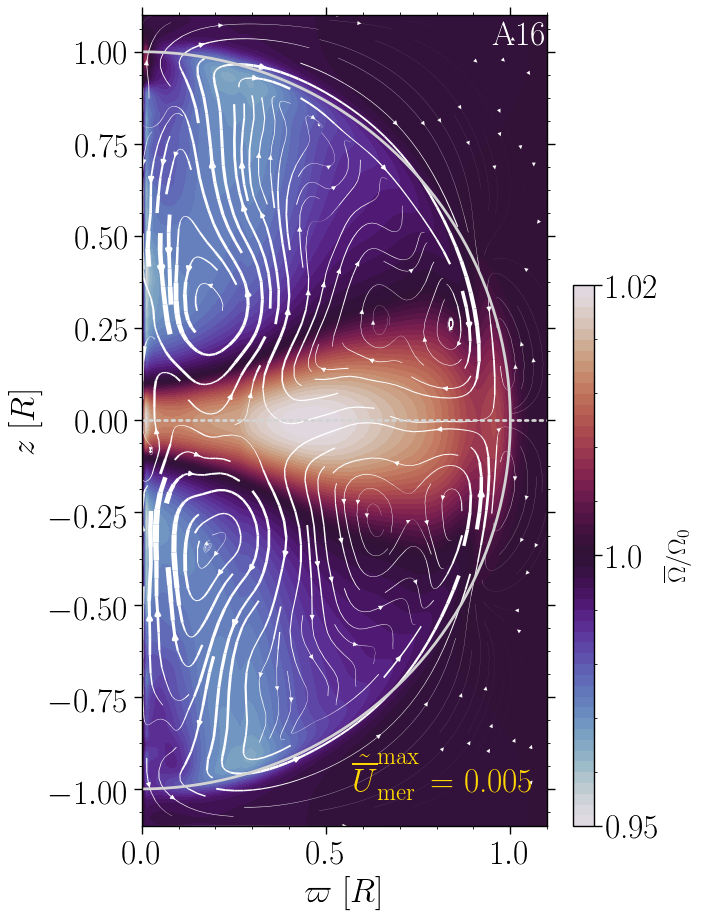}
\includegraphics[width=0.33\textwidth]{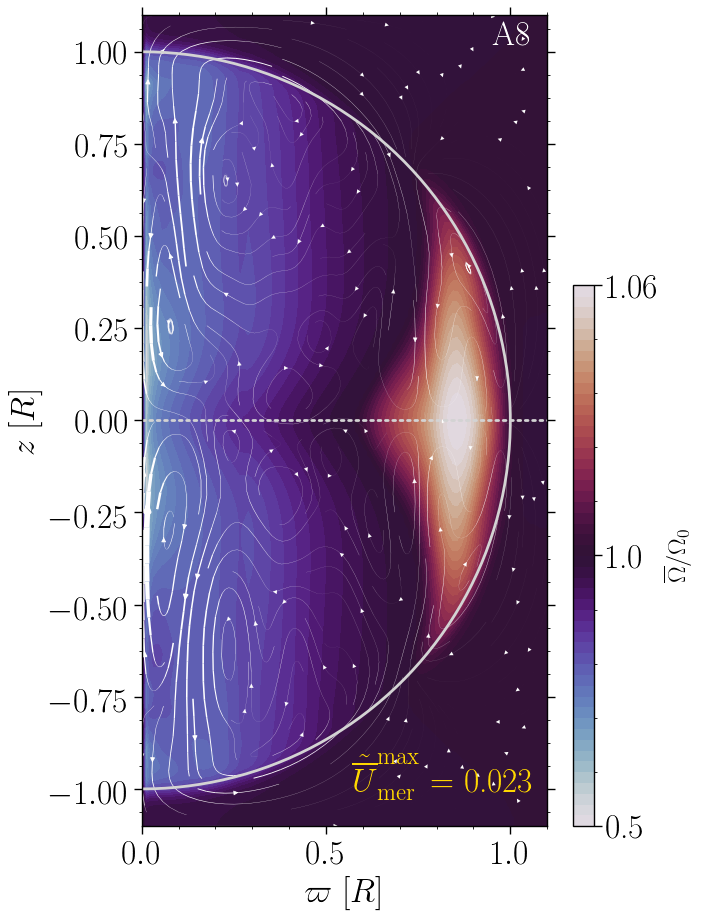}
\caption{{Normalized time-averaged mean angular velocity
    $\overline{\Omega}/\Omega_0$} for simulations A1 (left), A16 (middle), and 
  {A8} (right). {The colored contours indicate
  $\overline{\Omega}(\varpi,z)$. The streamlines indicate the mass
  flux due to meridional circulation. The amplitude of the meridional
  flow speed ($\tilde{\overline{U}}_{\rm mer}^{\rm max}$) is indicated in the
  lower right corner of each panel.} The surface is indicated by the
gray line while the equator is indicated by the gray dotted line.}
\label{fig:angular}
\end{figure*}

Values of $\Delta_{\Omega}^{(r)}$ and
$\Delta_{\Omega}^{(\overline{\theta})}$ are listed in
Table~\ref{tab:amplitude}. Positive values of $\Delta_{\Omega}^{(r)}$
indicate solar-like differential rotation. We find that
$\Delta_{\Omega}^{(r)}$ has a tendency to decrease with increasing
{$\rm Re_M$, which is equivalent to an increasing $\rm{Pr_M}$}
(second column of Table~\ref{tab:amplitude}). {Figure
  \ref{fig:deltaomr_rem} shows $\Delta_{\Omega}^{(r)}$ as a function
  of $\rm{Re_M}$ for sets A (circles), B (squares) and C (triangles) confirming the
  decreasing trend as a function of $\ReM$. The red circles show
  simulations without
dynamos (A8 and A12) and low magnetic Reynolds numbers, where
$\Delta_{\Omega}^{(r)}$ is higher. Such} reduction or quenching of
the differential rotation {by magnetic fields especially at high
  $\ReM$ has been shown earlier by various} simulations, for example,
in \cite{BMT04}, \cite{schrinner2012dipole}, and
\cite{kapyla2017convection}. {The differential rotation
  profiles for three representative simulations A1, A16 and A8 are
  shown in Fig.~\ref{fig:angular}.} The rotation profile
in run A1 is solar-like; the profile is similar in the rest of
simulations with ${\rm Pr_M}\leq 2$. In {run}
A16 with ${\rm Pr_M}=10$, the amplitude of the latitudinal
differential rotation is positive whereas the amplitude of the
radial differential rotation is negative since the angular velocity
does not change considerably with depth {at the
  equator}. The middle panel of Fig.~\ref{fig:angular} {shows that
  the rotation rate at the equator is in fact higher than average
  almost everywhere and the negative value of $\Delta_{\Omega}^{(r)}$
  is due to the slower than average rotation only very near the
  surface. Therefore the differential rotation is solar-like}. Another
method to classify the rotation
profile (solar-like or anti-solar) is to use the mean rotation profile
{at} the equator, which, as indicated in
\cite{2023A&A...669A..98K},
can help prevent {erroneous conclusions}. Furthermore, the
profiles
in Fig.~\ref{fig:angular} are symmetric {with respect} to the
equator as in the
other simulations performed in this work.

{The global} Coriolis number ({see
  Eq.~\ref{eq:coriolis})} in the current simulations ranges from
       {3.1} to {13}. All of our runs show solar-like
       differential rotation, which is
consistent with \cite{kaepylae2021}, where the shift from anti-solar
to solar-like differential rotation occurs for Coriolis number between
0.7 and 2.  This is also consistent with simulations of spherical
shell convection by \cite{viviani2018transition}, which show that the
transition occurs around ${\rm Co} = 3$. (see Table 5 of their
work). More recently, \cite{2023A&A...669A..98K} found that the
transition from anti-solar to solar-{like} differential rotation depends on
the sub-grid scale Prandtl number (${\rm Pr_{SGS}}$), such that
solar-like differential rotation is more difficult to obtain at high
${\rm Pr_{SGS}}$ than at ${\rm Pr_{SGS}} \leq 1$. In this work, all
the simulations have ${\rm Pr_{SGS}} \leq 1$.

Simulations {A8, A12, and C2 do not have dynamos, and they are
  considered as kinematic cases}. The right
panel of Fig.~\ref{fig:angular} displays the rotation profile for
simulation {A8}. It demonstrates that a faster {than average}
angular velocity spans a
broader latitudinal range and a {narrower} radial range when
compared to
simulations with dynamo. This depicts the influence of a magnetic
field on differential rotation. In the {regime $\PrM < 2$
  of} our simulations, the {meridional flow is} composed
of multiple small cells, while in the {regime} $\PrM \geq 2$,
the pattern is composed of two to three {large} cells which
are symmetric with respect to the equator. {The maximum values of the normalized meridional flow
  amplitude, $\tilde{\overline{U}}_{\rm mer}^{\rm rms} =
(GM/R)^{-1/2}(\overline{U}_{\varpi}^2 + \overline{U}_z^2)^{1/2}$, in the cases shown in Fig.~\ref{fig:angular}} correspond
to
$\tilde{\overline{U}}_{\rm mer}^{\rm max}=$ {0.009, 0.005 and 0.023} for
simulations A1, A16 and {A8 }, respectively. The rms value of the
meridional velocity is given in the fifth column of Table
\ref{tab:amplitude}. In the simulations with no dynamo, A8, A12 and C2
({A8} in right panel in Fig. \ref{fig:angular}), the meridional
circulation also exhibits {similar} multiple patterns, {which are also
  symmetric with respect to the equator.}

%PJK: Introduced a new subsection
\subsubsection{Power spectra and kinetic helicity}

\begin{figure}[!htbp]
\includegraphics[width=\linewidth]{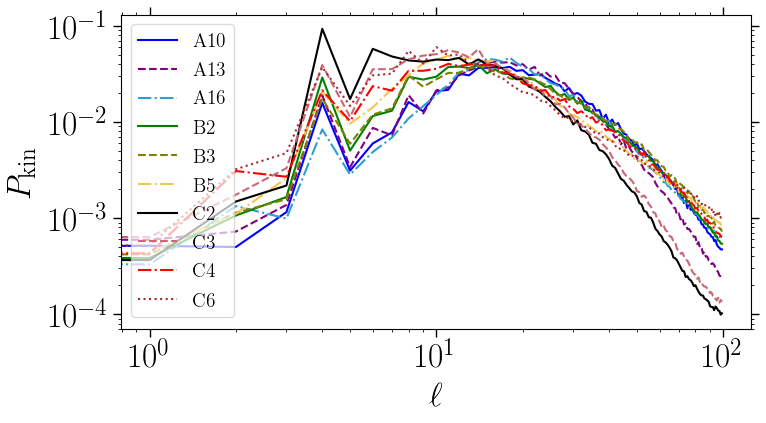}
\caption{Normalized convective power as a function of $\ell$ for
  simulations A10, A13, A16, B2, B3, B5, C2, C3, C4 and C6.}
\label{fig:convectivepower_l}
\end{figure}

{To characterize the convective flows we calculated the
  normalized kinetic energy power spectra
  \citep[e.g.][]{viviani2018transition,navarrete2022origin}} from
\begin{equation}
  P_{\rm kin} = \frac{E_{{\rm kin},\ell}}{\sum_\ell E_{{\rm kin},\ell}},
\end{equation}
{where $E_{{\rm kin},\ell}$ is the kinetic energy of the spherical
  harmonic degree $\ell$, that is calculated {from} the
  decomposition of the
%PJK: Why only radial? We can do the full power spectrum now, right?
  radial velocity field at the surface into spherical
  harmonics. Figure~\ref{fig:convectivepower_l} shows $P_{\rm kin}$ as
  a function of $\ell$ for
  selected simulations. For the simulations with lower rotation rates
  and large-scale dynamos, the convective power is slightly shifted
  towards lower $\ell$, with peaks between 16 and 20 for set A, 12 and 15
  for set B, and 7 and 15 for set C. In simulations with
  no dynamo, the peak is at considerably larger scales at
    $\ell=4$. This demonstrates the suppression of large-scale convective
    flows by magnetic fields. This is reminiscent to results from
    recent solar-like simulations that suggest that suppression of
    large-scale convection may be important to maintain a solar-like
    rotation profile in the Sun
    \citep[e.g.][]{2022ApJ...933..199H,2023A&A...669A..98K}. Furthermore,
    the large-scale convective amplitudes are also in general higher
    in cases with slower rotation in accordance with linear theory
    \citep{Ch61} and various earlier simulations
    \citep{FH16b,viviani2018transition, navarrete2022origin}}.

\begin{figure}[!htbp]
\centering
\includegraphics[width=0.7\linewidth]{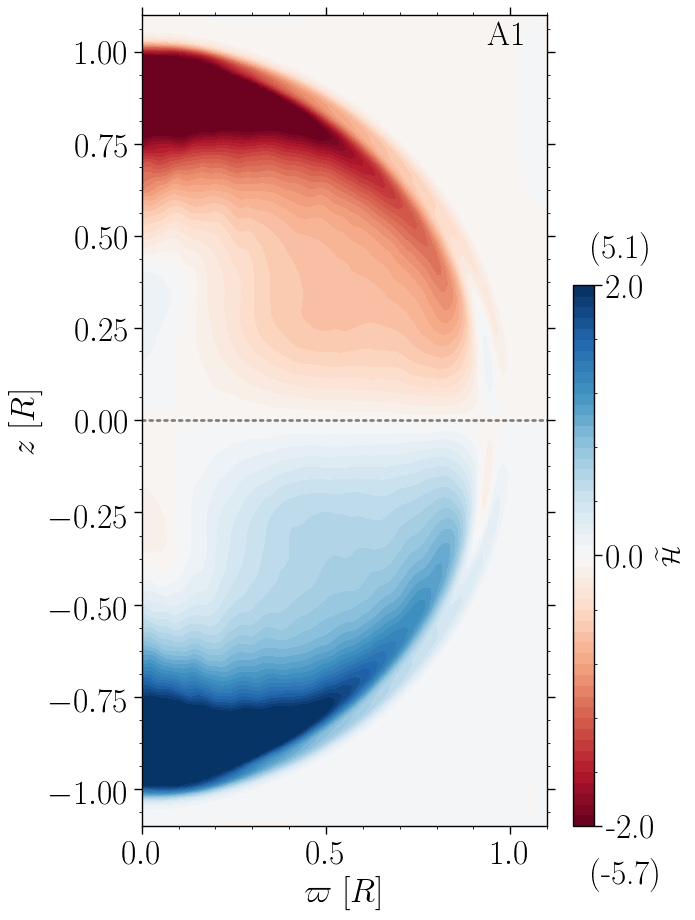}
\caption{Azimuthally averaged normali{z}ed kinetic helicity,
  ${\widetilde{\mathcal{H}} = \mathcal{H}(\varpi, z)/u_{\rm rms}
    \omega_{\rm rms}}$, for simulation A1 with $\Co = 9.2$ and
  $\PrM = 0.2$.}
\label{fig:kinetic_helicity}
\end{figure}

The kinetic helicity, defined as $\mathcal{H} = \overline{\bm{\omega}
  \bm\cdot \bm{u}}$ is an important component in the operation of the
dynamo. It is a proxy of the $\alpha$-effect, which is
responsible for producing {poloidal fields from toroidal fields
  (and vice versa)} by rising {or descending} and twisting
convective eddies
\citep{parker1955hydromagnetic,1966ZNatA..21..369S}. In all of our
simulations the kinetic
helicity is negative (positive) in the northern (southern) hemisphere,
as is shown Fig. \ref{fig:kinetic_helicity} for run A1. This,
combined with a solar-like differential rotation, suggests that an
$\alpha\Omega$ dynamo is operating, in which {case} the direction of
propagation of the dynamo waves is poleward
\citep{parker1955hydromagnetic, yoshimura1975solar}. This is
%PJK: better not refer to figures here
consistent with our findings, {which will be discussed in more
  detail in Sect.~\ref{sec:dynamo}}.

%PJK: The discussion of the fluxes fits poorly under the heading "Flow properties''
%PJK: Introduced a new subsection
\subsubsection{Convective energy transport}
\begin{figure}[!htbp]
\includegraphics[width=\linewidth]{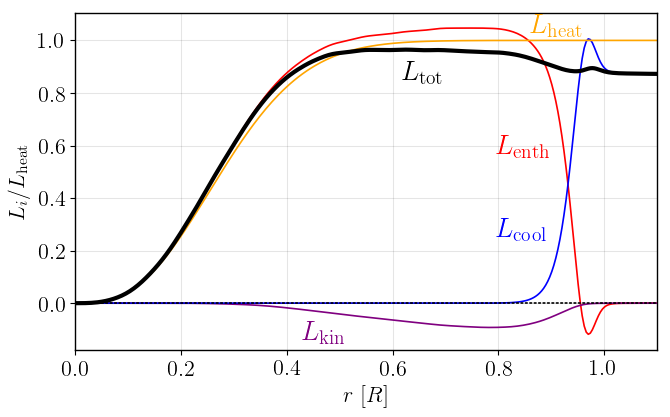}
\caption{Luminosity profiles of kinetic energy (purple), cooling (blue), heating (orange) and enthalpy (red) fluxes of run C4. }
\label{fig:lum_profile}
\end{figure}

{The luminosities corresponding to radiative, enthalpy, kinetic
  energy, cooling and heating fluxes according to Eqs.(31) to (36) of
  \cite{kaepylae2021} are shown in Fig.~\ref{fig:lum_profile} for run
  C4. The enthalpy and kinetic energy fluxes dominate almost
  everywhere, except near the
  surface where the cooling becomes important. This is similar to the
  results of \citet{brown2020single} and of the rotating runs of
  \citet{kaepylae2021}.
The total flux reaches somewhat less than 90 per cent of the
luminostiy from the heating near the surface. A possible reason for
this discrepancy is a non-negligible contribution from the SGS flux.}

\subsection{Dynamo variation}
\label{sec:dynamo}

\begin{figure*}[!htbp]
\begin{multicols}{3}
\begin{subfigure}{0.33\textwidth}
\includegraphics[width=\linewidth]{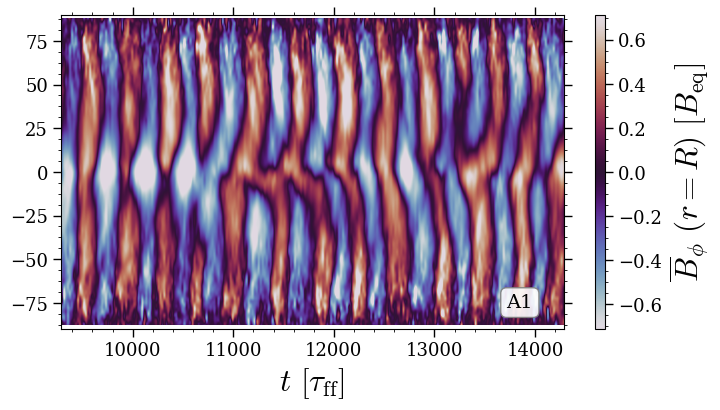}
\end{subfigure}
\begin{subfigure}{0.33\textwidth}
\includegraphics[width=\linewidth]{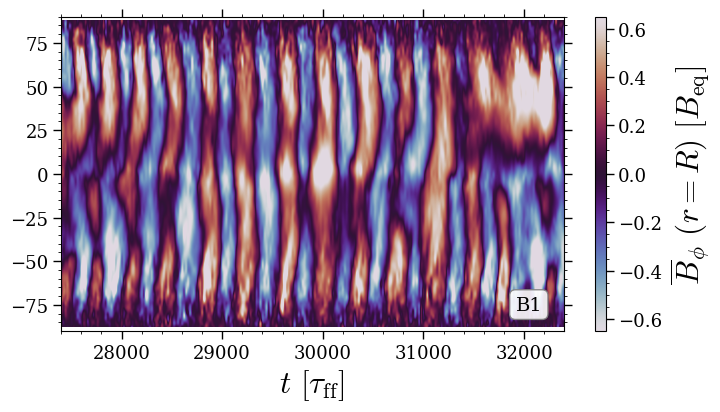}
\end{subfigure}
\begin{subfigure}{0.33\textwidth}
\includegraphics[width=\linewidth]{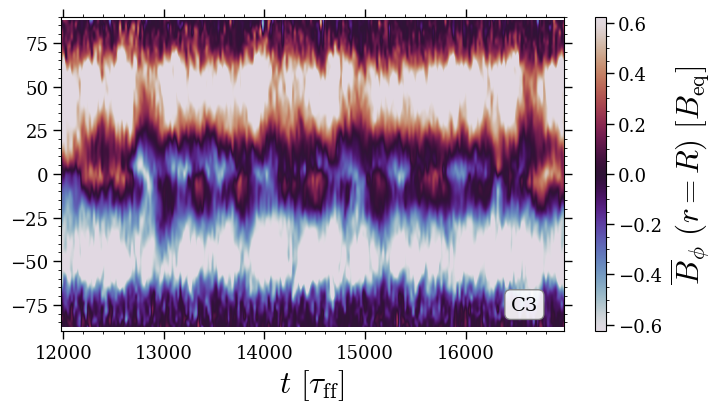}
\end{subfigure}
\begin{subfigure}{0.33\textwidth}
\includegraphics[width=\linewidth]{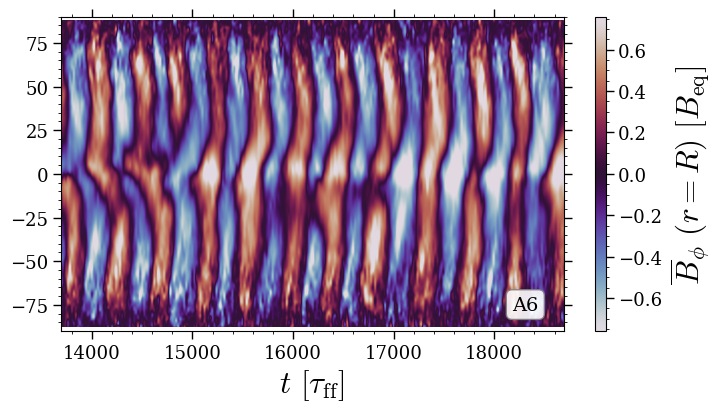}
\end{subfigure}
\begin{subfigure}{0.33\textwidth}
\includegraphics[width=\linewidth]{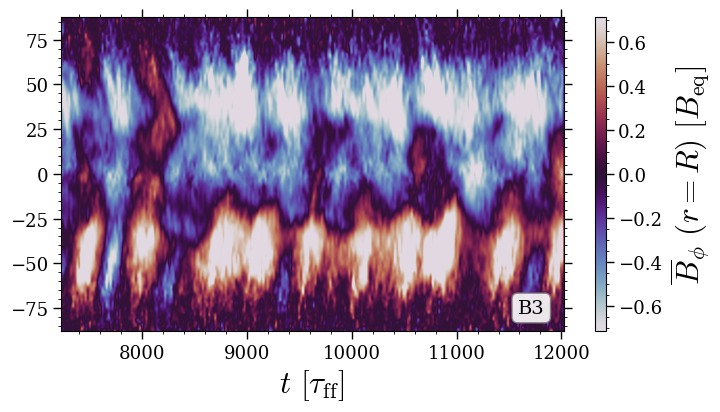}
\end{subfigure}
\begin{subfigure}{0.33\textwidth}
\includegraphics[width=\linewidth]{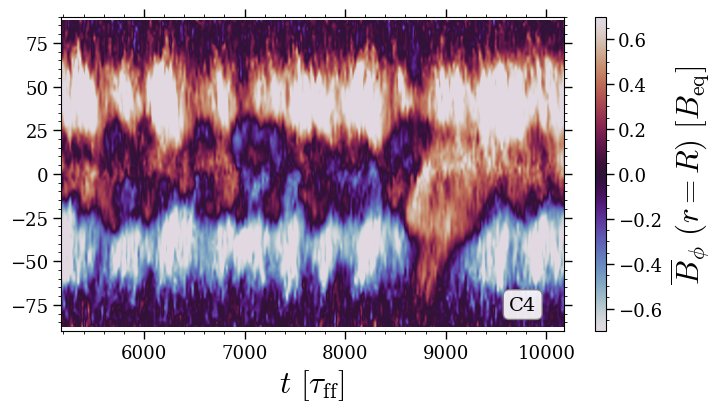}
\end{subfigure}
\begin{subfigure}{0.33\textwidth}
\includegraphics[width=\linewidth]{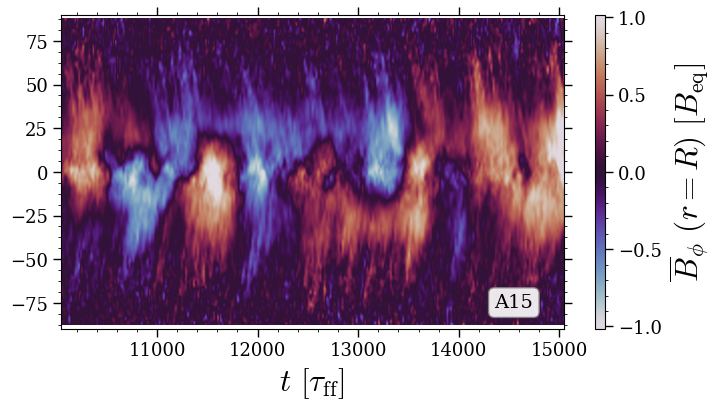}
\end{subfigure}
\begin{subfigure}{0.33\textwidth}
\includegraphics[width=\linewidth]{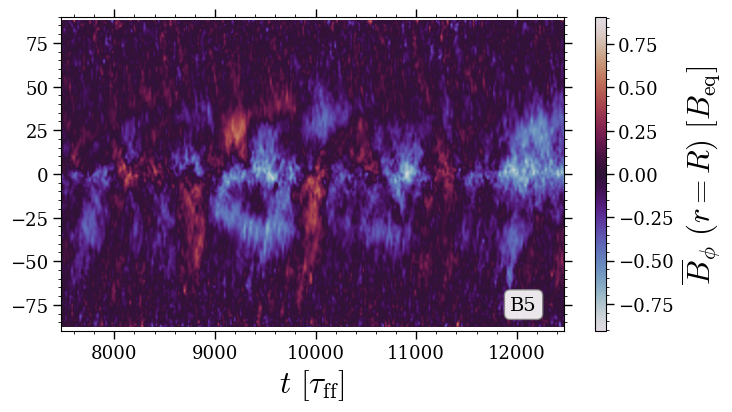}
\end{subfigure}
\begin{subfigure}{0.33\textwidth}
\includegraphics[width=\linewidth]{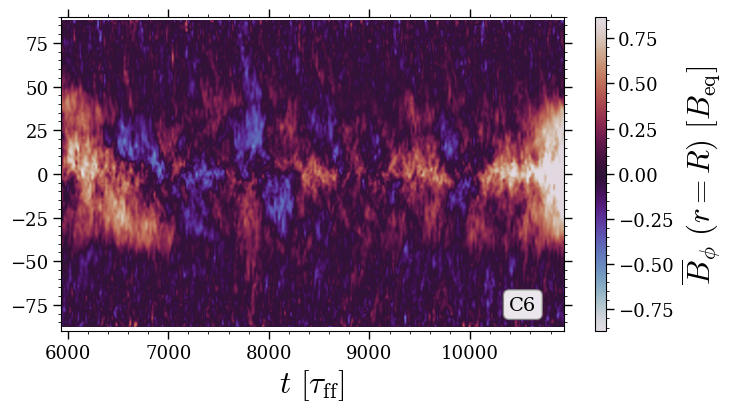}
\end{subfigure}
\end{multicols}
\caption{Azimuthally averaged toroidal magnetic field near the surface of the star {as a function of time}. The name of the simulation is indicated bottom right of each panel.}
\label{fig:Bpphi_r=R}
\end{figure*}
\begin{figure*}[!htbp]
\begin{subfigure}{0.33\textwidth}
\includegraphics[width=\linewidth]{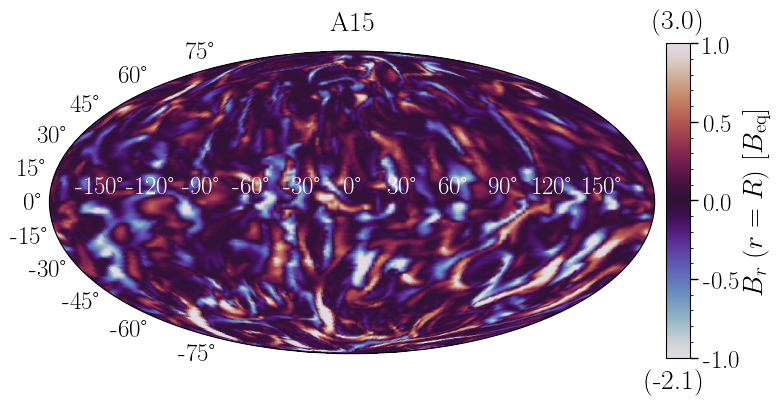}
\end{subfigure}
\begin{subfigure}{0.33\textwidth}
\includegraphics[width=\linewidth]{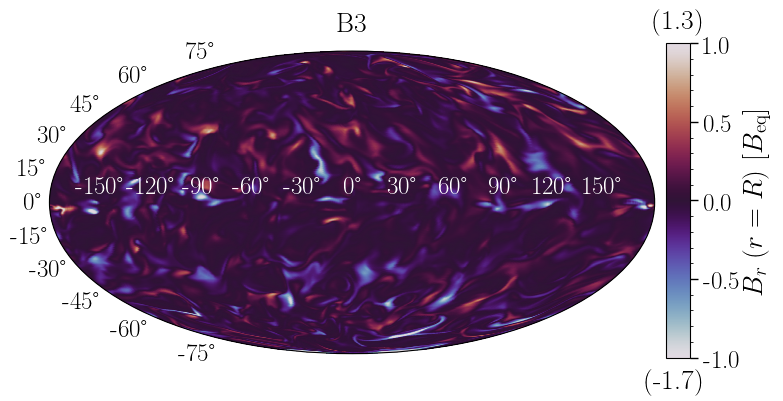}
\end{subfigure}
\begin{subfigure}{0.33\textwidth}
\includegraphics[width=\linewidth]{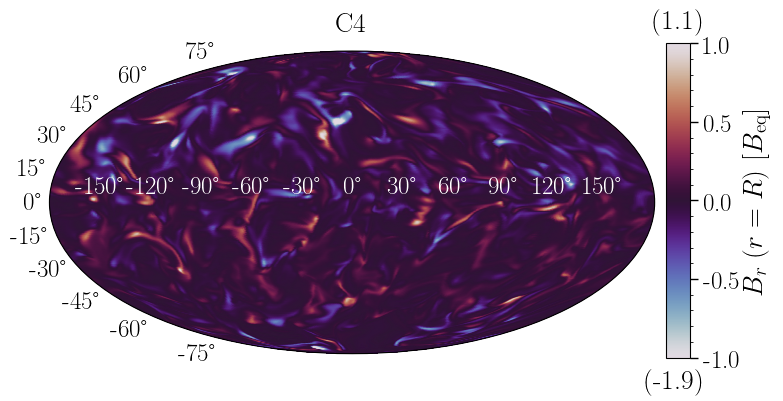}
\end{subfigure}
\caption{Mollweide projection of the radial magnetic
  field $B_r$ for {runs
    A15, B3 and C4 with $\ReM \approx 300$ and different rotation
    rates.}}
\label{fig:Br_1}
\end{figure*}
\begin{figure}[!htbp]
\includegraphics[width=\linewidth]{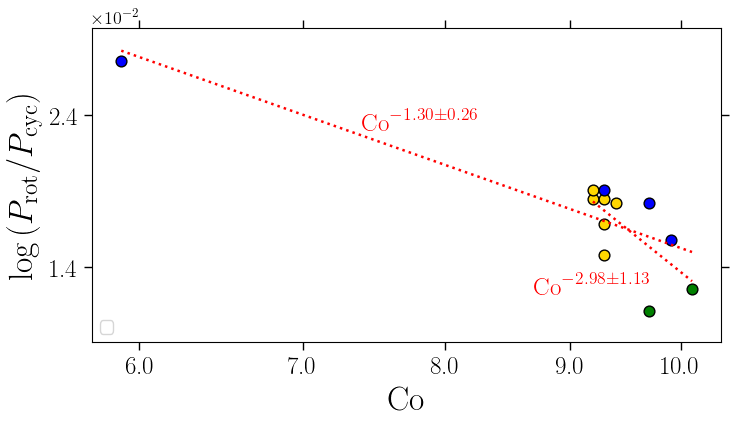}
\caption{Rotation period normalized to the cycle period as a function of the global Coriolis number. Green circles are $\ReM \sim 100$, blue for $ 70 < \ReM < 85$, and yellow is for $\ReM \leq 55$.}
\label{fig:prot_pcyc_vs_co}
\end{figure}

As shown in Table~\ref{tab:summary_table}, the main differences
between the simulations are the input parameters $\PrM$ and
{the rotation rate}. Next, we present the effects of varying these
parameters on
%PJK
the large-scale magnetic field.

\subsubsection{Dependence on rotation}

We have explored simulations with fixed $\PrM$ and varying $\Prot$
with values between 43 to 90 days. {These values were determined
  using the conversion method outlined in Appendix~A of
  \cite{kaepylae2021}}. In
order to compare the large-scale magnetic field at the three rotation
rates used here, we chose {runs with comparable magnetic Reynolds
  numbers and different rotation rates. }

{Three representative runs with $\ReM < 100$ from each set are A6
  with $\ReM = 75$, B1 with $\ReM = 84$ and C3 with $\ReM = 67$. A6
  shows cycles in the azimuthally averaged toroidal magnetic field,
%PJK: the panels come in a confusing order now...
  $\overline{B}_{\phi}(R,\theta,t)$, as shown in the middle top panel
  of Figure \ref{fig:Bpphi_r=R}. The cycles were computed using the
  empirical mode decomposition with the \textit{libeemd} library
  \citep{luukko2016introducing}, as in 
  \citet{kapyla2022solar}. To determine the periods we use
  $\overline{B}_{\phi}(R,\theta,t)$, from the range
  $-60^{\circ}<\theta<60^{\circ}$. The cycle is determined by taking
  the mode
  with the largest energy, and counting the period from the zero
  crossings of that mode.}

{Left middle panel of Fig.~\ref{fig:Bpphi_r=R} shows
  $\overline{B}_{\phi}(R,\theta,t)$ for run B1, which also exhibits
  cycles. The reversals are periodic for
  most of the run, and it also shows longer term modulation in the
  northern hemisphere toward the end of the run. The left bottom panel
  of
  Fig.~\ref{fig:Bpphi_r=R} is for run C3. Unlike the runs just
  mentioned, C3 does not exhibit cyclic reversals. However, it does
  reveal the presence of a dipolar field, with a positive (negative)
  polarity in the northern (southern) hemisphere. At similar
  values of magnetic Reynolds number, the third column of Table
  \ref{tab:summary_table} indicates a slight reduction in $B_{\rm
    rms}$ at lower rotation rates}.

{Three representative runs with higher magnetic Reynolds number
  ($\ReM \approx 300$) and different rotation rates are A15 with $\ReM =
  300$, B3 with $\ReM = 315$ and C4 with $\ReM = 337$. The right top
  panel of Fig.~\ref{fig:Bpphi_r=R} shows
  $\overline{B}_{\phi}(R,\theta,t)$ of run A15. This run has irregular
  reversals with the field mainly distributed from mid-latitudes
  ($\pm 45\degr$) to the equator. Near the poles the field is
  quasi-stationary. The middle center panel of Figure
  \ref{fig:Bpphi_r=R} is for B3, where a dipole with a few random
  reversals is visible with a predominantly negative (positive)
  polarity at the northern (southern)
  hemisphere. $\overline{B}_{\phi}(R,\theta,t)$ of C4 is shown in the
  center bottom panel of Fig.~\ref{fig:Bpphi_r=R}, where a
  predominantly positive (negative) polarity. Mollweide projections of
  the radial magnetic field at the surface of runs A15, B3 and C4 are shown in  Fig.~\ref{fig:Br_1}, where the field is less
  intense for the runs with lower rotation. In this sense, the $B_{\rm
    rms}$ decreases with decreasing the rotation rate from A15 to B3,
  while B3 and C4 have similar values. We find that in general the
  saturation level of the magnetic field increases with $\ReM$.
This behavior is likely related to the presence of a small-scale
dynamo that produces magnetic fields at spatial scales that are of the
same order of magnitude as that of the turbulence. While this was not the focus of our current study this remains an important are for future research.}

{In Fig.~\ref{fig:prot_pcyc_vs_co} we show the ratio of the
rotation period to cycle period as a function of the global Coriolis
number. We find that $P_{\rm rot}/P_{\rm cyc} \propto
    \Co^{\beta}$ with $\beta = -1.30 \pm 0.26$. When considering the data points on the right of the figure, we find that $\beta = -2.98 \pm 1.13$. The uncertainty in the slope indicates that we need to take these results with caution. Nevertheless, earlier studies have also
found $\beta < 0$, for example, \cite{strugarek2017reconciling,
  strugarek2018sensitivity, warnecke2018dynamo, viviani2018transition}
with global simulations of solar-like stars. Even when the domain of
those simulations differs from the one presented here, the similarity
in the relationship between the cycle period and the Coriolis number
implies a likeness in the dynamo processes of solar-like and fully
convective stars. Nevertheless, the negative slope found here differs
from the positive slopes for the inactive and active branches from
observations \cite{brandenburg1998time,
  brandenburg2017evolution}. However, also some simulations show
$\beta \gtrsim 0$ \citep{2019ApJ...880....6G,kapyla2022solar}, but the
cause of such behavior is currently unclear.}

\subsubsection{Dependence on {magnetic Reynolds and Prandtl numbers}}

Magnetic Prandtl numbers from $0.1$ to $10$ were used in the
simulations. For all the current runs, the magnetic field is
predominantly axisymmetric.
{When converted to physical units, the azimuthally averaged toroidal magnetic field reaches strengths ranging from 10 to 16 kG in our models.} These values are higher
than those of the reported observations {which are up to a few
  kG \citep[e.g.][]{kochukhov2021magnetic}.} Set A
{has cycles for
  ${\rm Pr_M} \leq 2$ with periods ranging from 309 to 471 freefall times, which correspond to 6.3 to 9.6 years, when
  considering the same time conversion factor used by
%PJK: need to use non-dimensional units here
  \citet{kaepylae2021}. Run B1 also shows cycles with a period of 274 freefall times. Table~\ref{tab:energies} lists the values of $\nu$, $ \eta$,
%  the length of the cycles (if applicable) together with the
%PJK: cycle periods
  the cycle periods (if applicable) together with the
  corresponding standard deviation for all the simulations presented
  here. We found that the calculated length of the cycle periods of
  the runs of set A has a very slight increase when increasing the
  magnetic Reynolds number as ${\Pcyc} \propto \ReM^{\alpha}$
  with $\alpha = 0.25\pm 0.14$. Additionally, when considering the
  runs with similar ${\ReM}$ and different ${\rm Pr_M}$, we
  found that the cycle period is virtually independent of ${\rm Pr_M}$
  in the parameter regime explored here. }

{The azimuthally averaged toroidal magnetic field
  $\overline{B}_{\phi}(R,\theta,t)$ is shown in Figure
  \ref{fig:Bpphi_r=R} for a set of representative runs. The top panels
  are for three runs
  of set A, which have the same rotation period and increasing
  ${\ReM}$ from left to right. The top left panel is for run A1 with
  $\ReM = 55$, with $\Pcyc = 320 \pm 10$ freefall times. The
  top center panel is for run A6 with $\ReM = 75$ and $\Pcyc = 326 \pm 11$ freefall times. In these cases the field is distributed in
  latitudes $|\theta| \lesssim 80^{\circ}$. Simulations with higher
  values of ${\rm Pr_M}$ and $\ReM$, such as run A15 with $\ReM =
  300$, result in the loss of cycles and the emergence of irregular
  solutions. Similar irregularity of dynamo solutions has previously
  been observed in simulations with high $\ReM$
  \citep[e.g.][]{kapyla2017convection}, but the exact mechanims is
  still unknown. In this case, the field is distributed
  at latitudes $|\theta| \lesssim 50$ and also exhibits
  quasi-stationary solutions near the poles.}
 
\begin{figure}[!htbp]
\begin{subfigure}{0.5\textwidth}
\includegraphics[width=\linewidth]{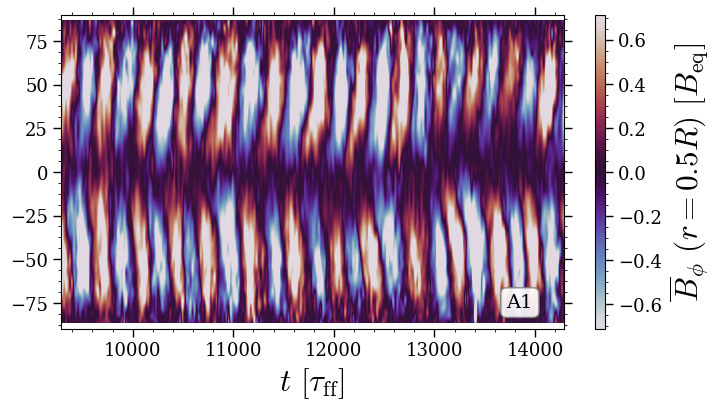}
\end{subfigure}
\begin{subfigure}{0.5\textwidth}
\includegraphics[width=\linewidth]{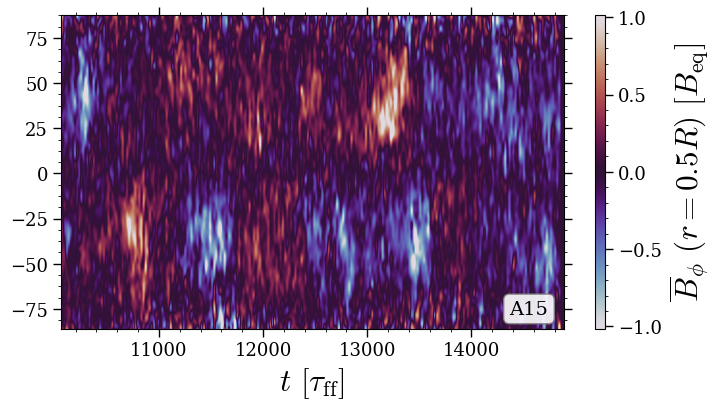}
\end{subfigure}
\caption{Azimutally averaged toroidal magnetic field at $r=0.5$ for simulations A1 (top) and A15 (bottom) with ${\ReM} = 55$ and ${\ReM} = 300$, respectively.}
\label{fig:Bpphi_radii}
\end{figure}

{The} polarity {of the field}
changes from {the surface} to $r=0.5\, R$. {Figure
  \ref{fig:Bpphi_radii} shows $\overline{B}_{\phi}$ at {$r=0.5R$
    for runs} A1 and A15. In simulations with cycles, such as A1,
  the cycles are visible throughout the convection zone. However, for
  runs with higher} magnetic Prandtl number ($\rm Pr_M > 2$), {such as A15} the azimuthally averaged toroidal
magnetic field changes with depth {and shows} less clear magnetic
structures in the deeper layers.

\begin{figure*}[!htbp]
%\begin{multicols}{3}
\begin{subfigure}{0.33\textwidth}
\includegraphics[width=\linewidth]{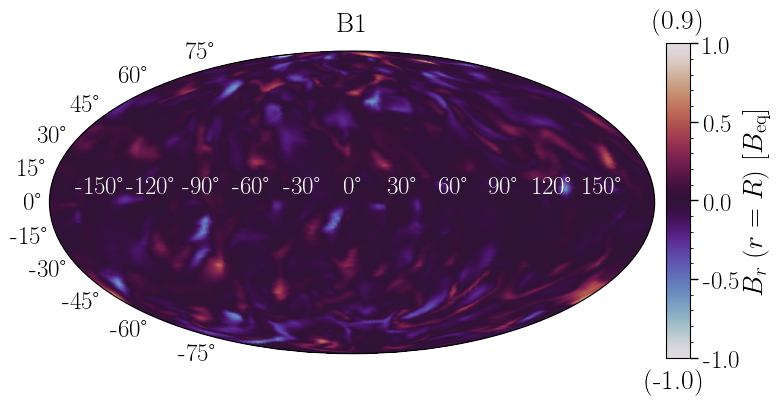}
\end{subfigure}
\begin{subfigure}{0.33\textwidth}
\includegraphics[width=\linewidth]{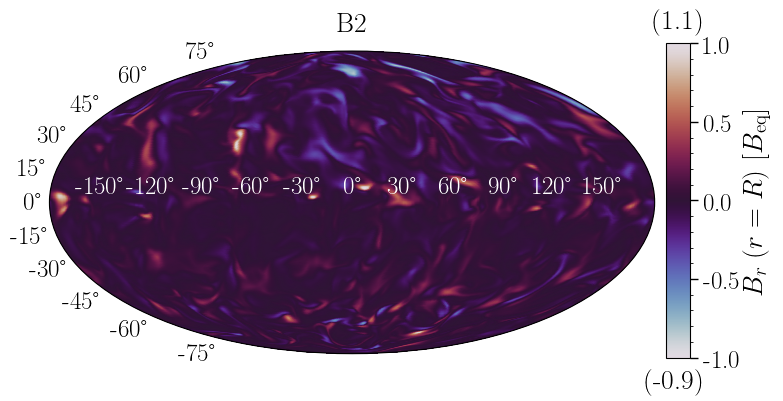}
\end{subfigure}
\begin{subfigure}{0.33\textwidth}
\includegraphics[width=\linewidth]{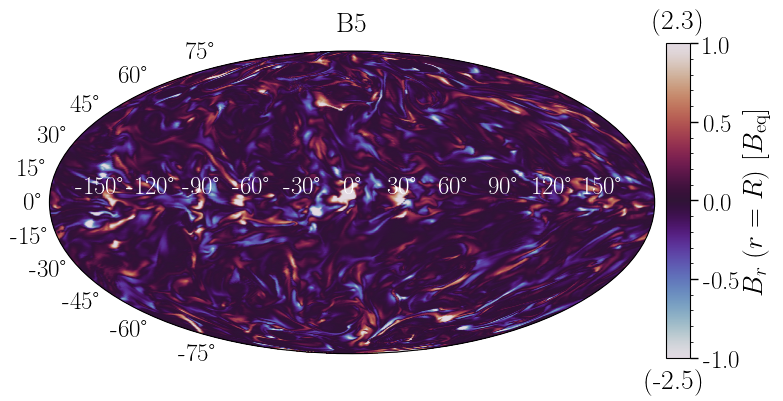}
\end{subfigure}
%\end{multicols}
\caption{Mollweide projection of the radial magnetic
  field $B_r$ for runs with the same rotation rate and different $\ReM$,
    B1 with $\ReM = 84$, B2 with $\ReM = 168$, and B5
    with $\ReM = 1360$.}
\label{fig:Br_2}
\end{figure*}

{The second row of Fig.~\ref{fig:Bpphi_r=R} shows three
  runs from set B which have the same rotation rate and increasing
  $\ReM$. The left panel is for run B1 with $\ReM = 84$, which
  exhibits a cycle with $\Pcyc = 274 \,\pm\,38$ freefall times, as well as
  longer reversals or disappearing cycles toward the end of the
  run. In this run the field is distributed at
  latitudes $|\theta| \lesssim 80^{\circ}$. The center panel is for
  run B3 with $\ReM = 315$, which exhibits an irregular solution with
  few polarity reversals and predominantly quasi-static fields. In
  this
  case, the field spans slightly less latitudinally, distributed at
  latitudes $|\theta| \lesssim 75^{\circ}$. The right center panel
  Fig.~\ref{fig:Bpphi_r=R} shows run B5, which has the highest ${\rm
    Pr_M}$ and $\ReM$ in this set with ${\rm Pr_M}=10$ and
  $\ReM=1360$. The
  field is more concentrated toward the equator (at latitudes $|\theta| \lesssim 50^{\circ}$) with seemingly irregular reversals.
%PJK: this cannot really be seen
  Similar to run A15, B5 also has a quasi-stationary solution
  near the poles. The third row of Fig.~\ref{fig:Bpphi_r=R}
  are for runs of set C, which have the slowest rotation in the
  present work, with increasing $\ReM$ from left to
  right. Runs C3 with $\ReM = 67$ and C4 with $\ReM = 337$ show a
  predominantly quasi-static dipolar field, which spans latitudes
  $|\theta|
%PJK
  \lesssim 75^{\circ}$. A similar dipolar field was reported by
  \cite{moutou2017spirou} for the fully convective and slow-rotating M
  dwarf GJ 1289. However, in our models the toroidal magnetic
  energy is dominant (see Table \ref{tab:energies}), whereas the
%PJK
  large-scale magnetic field of GJ 1289 is purely poloidal. Run C6
  with $\ReM=1419$ has a field concentrated near the equator and the
  large-scale structures are less clear than in C3 and C4.}

The radial magnetic field, $B_r$, also varies as a
function of $\ReM$. Mollweide projections of the radial magnetic field
for {runs B1, B2 and B5, with
  increasing $\ReM$ from left to right are} presented in Figure 
\ref{fig:Br_2}. {The main differences here are
  the structure and maximum values of the magnetic field strength. The
  size of the structures in runs B1 and B2 are similar, but the
  strength of the field is slightly higher in B2. Run B5 has smaller
  field structures than in the previous cases, and the magnetic field
  strength is higher.}

Table \ref{tab:energies} lists the energy densities of the
simulations. The {total magnetic energy $E_{\rm mag}$ is a
  significant fraction of the kinetic energy density in all of the
  runs with dynamos, sometimes also exceeding it.} 
 One may expect that $E_{\rm mag}$ grows with increasing ${\rm Re_M}$ ($\rm Pr_M$), as
found in other works (e.g., \citealt{kapyla2017convection}).  In this
regard, there is no discernible trend in the simulations shown here in
terms of the variation of $E_{\rm mag}$. Since the kinetic energy
density, $E_{\rm kin}$, decreases with increasing $\rm Pr_M$ the ratio
$E_{\rm mag}/E_{\rm kin}$ grows. The decrease of the kinetic energy
can be explained because at large ${\rm Pr_M}$, it is converted into
magnetic energy more efficiently.

\begin{table*}[!htbp]
\centering
\begin{tabular}{ccccccccccc}
\hline
%PJK: nu, eta, and cycle periods should be non-dimensionalized
Sim  & $\nu$                     & $\eta$                     & Cycles  & ${\boldsymbol{\sigma}}$ & ${\Tilde{E}_{\rm mag}}$ & $E_{\rm mag}^{\rm tor}/E_{\rm mag}$ & ${E_{\rm mag}^{\rm pol}/E_{\rm mag}}$ & ${\Tilde{E}_{\rm kin}^{\rm DR}} $ & ${\Tilde{E}_{\rm kin}^{\rm MC}}$ \\ \hline
A1   & $1.0\cdot10^{-5}$   & $1.0 \cdot 10^{-4}$  & 320 & 10             & 0.86                    & 0.23                                & 0.07                                  & 0.23                              & 0.01                             \\
A2   & $6.9 \cdot 10^{-6}$ & $6.9 \cdot 10^{-5}$  & 309 & 7              & 0.89                    & 0.20                                & 0.06                                  & 0.27                              & 0.02                             \\
A3   & $2.0 \cdot10^{-5}$  & $1.0 \cdot 10^{-4}$  & 350 & 13             & 0.85                    & 0.24                                & 0.06                                  & 0.24                              & 0.06                             \\
A4   & $5.0 \cdot 10^{-5}$ & $1.0 \cdot 10^{-4}$  & 320 & 33             & 0.64                    & 0.20                                & 0.06                                  & 0.31                              & 0.02                             \\
A5   & $7.0 \cdot 10^{-5}$ & $1.0 \cdot 10^{-4}$  & 310 & 72             & 0.41                    & 0.24                                & 0.06                                  & 0.34                              & 0.02                             \\
A6   & $4.9 \cdot 10^{-5}$ & $6.9 \cdot 10^{-5}$  & 326 & 11             & 0.84                    & 0.21                                & 0.05                                  & 0.31                              & 0.02                             \\
A7   & $5.0 \cdot 10^{-5}$ & $1.0 \cdot 10^{-4}$  & 324 & 17             & 0.68                    & 0.25                                & 0.06                                  & 0.31                              & 0.02                             \\
A8*  & $7.0 \cdot 10^{-5}$ & $3.5 \cdot 10^{-4}$  & -   & -              & -                       & -                                   & -                                     & 0.55                              & 0.01                             \\
A9   & $7.0 \cdot 10^{-5}$ & $1.4 \cdot 10^{-4}$  & 387 & 63             & 0.39                    & 0.28                                & 0.07                                  & 0.32                              & 0.02                             \\
A10  & $5.0 \cdot 10^{-5}$ & $5.0 \cdot 10^{-5}$  & 471 & 84             & 0.77                    & 0.16                                & 0.04                                  & 0.32                              & 0.02                             \\
A11  & $7.0 \cdot 10^{-5}$ & $7.0 \cdot 10^{-5}$  & 368 & 23             & 0.78                    & 0.20                                & 0.05                                  & 0.30                              & 0.02                             \\
A12* & $1.0 \cdot 10^{-4}$ & $1.0 \cdot 10^{-4}$  & -   & -              & -                       & -                                   & -                                     & 0.54                              & 0.02                             \\
A13  & $1.0 \cdot 10^{-4}$ & $5.0 \cdot 10^{-5}$  & 437 & 13             & 1.44                    & 0.15                                & 0.04                                  & 0.32                              & 0.02                             \\
A14  & $1.0 \cdot 10^{-4}$ & $2.0 \cdot 10^{-5}$  & -   & -              & 1.68                    & $0.08$                               & 0.03                                  & 0.16                              & 0.02                             \\
A15  & $1.0 \cdot 10^{-4}$ & $7.1 \cdot 10^{-5}$  & -   & -              & 1.46                    & $0.06$                               & 0.02                                  & 0.14                              & 0.02                             \\
A16  & $1.0 \cdot 10^{-4}$ & $1.0\cdot 10^{-5}$   & -   & -              & 1.74                    & 0.03                                & 0.02                                  & 0.10                              & 0.01                             \\ \hline
B1   & $3.6\cdot 10^{-5}$  & $7.2\cdot 10^{-5}$   & 274 & 38             & 0.45                    & $0.20$                              & 0.04                                  & 0.33                              & 0.02                             \\
B2   & $3.6\cdot 10^{-5}$  & $3.6\cdot 10^{-5}$   & -   & -              & 0.53                    & 0.16                                & 0.03                                  & 0.38                              & 0.02                             \\
B3   & $3.6\cdot 10^{-5}$  & $1.8\cdot 10^{-5}$   & -   & -              & 0.80                    & 0.11                                & 0.03                                  & 0.30                              & 0.02                             \\
B4   & $3.6\cdot 10^{-5}$  & $7.2\cdot 10^{-6}$   & -   & -              & 1.21                    & 0.06                                & 0.02                                  & 0.22                              & 0.02                             \\
B5   & $3.6\cdot 10^{-5}$  & $3.6\cdot 10^{-6}$   & -   & -              & 1.13                    & 0.03                                & 0.02                                  & 0.014                             & 0.01                             \\ \hline
C1   & $2.5 \cdot 10^{-5}$ & $2.5 \cdot 10^{-5}$  & -   & -              & 0.67                    & 0.17                                & 0.03                                  & 0.31                              & 0.03                             \\
C2*  & $1.0 \cdot 10^{-4}$ & $2.0 \cdot 10^{-4}$  & -   & -              & -                       & -                                   & -                                     & 0.14                              & 0.12                             \\
C3   & $1.0 \cdot 10^{-4}$ & $1.0 \cdot 10^{-4}$  & -   & -              & 0.28                    & 0.34                                & 0.03                                  & 0.44                              & 0.02                             \\
C4   & $3.6 \cdot 10^{-5}$ & $1.8 \cdot 10^{-5}$  & -   & -              & 0.81                    & 0.12                                & 0.03                                  & 0.25                              & 0.03                             \\
C5   & $3.6 \cdot 10^{-5}$ & $7.2 \cdot 10^{-6}$  & -   & -              & 1.72                    & 0.10                                & 0.03                                  & 0.19                              & 0.03                             \\
C6   & $3.6 \cdot 10^{-5}$ & $3.6 \cdot 10^{-6}$  & -   & -              & 1.06                    & 0.04                                & 0.02                                  & 0.16                              & 0.02                             \\ \hline
\end{tabular}
\caption{Columns from left to right indicate {normalized} kinematic viscosity $\tilde{\nu} = (RGM)^{-1/2} \nu$, {normalized} magnetic diffusivity {\bf ${\tilde{\eta}}= (RGM)^{-1/2} \eta$}, cycle periods, {and their
    standard deviations, both in terms of $t_{\rm ff}$, and energy densities} if applicable. The
  magnetic energy density is $E_{\rm mag} =
  \langle{\bm{B}^2/2\mu_0}\rangle$, {where} the brackets indicate
  volume {and time}
  average within the radius of the star. The kinetic energy density is
  $E_{\rm kin} = \frac{1}{2} \langle \rho {\bm U}^2 \rangle$. The
  energy density for the azimuthally averaged toroidal and poloidal
  fields are given by $E_{\rm mag}^{\rm tor} =
  \langle \overline{B}^2_{\phi}/2\mu_0 \rangle$, and $E_{\rm mag}^{\rm
    pol} = (\langle
  \overline{B}^2_{\varpi} + \overline{B}^2_{z}\rangle) /2\mu_0$,
  respectively. The energy density for the differential rotation and
  meridional circulation are given by ${E_{\rm kin}^{\rm DR}} =
  \frac{1}{2} \langle \rho { \overline{U}}^2_{\phi} \rangle$, and
  $E_{\rm kin}^{\rm MC} = \frac{1}{2}\langle \rho(
  \overline{U}_{\varpi}^2 + \overline{U}_{z}^2) \rangle$,
  respectively. {Tildes over energies refer
    to normalization by $E_{\rm kin}$}. Asterisks indicate runs with
  no dynamo.}
\label{tab:energies}
\end{table*}
 
\begin{figure}[h!]
\centering
\includegraphics[width=\linewidth]{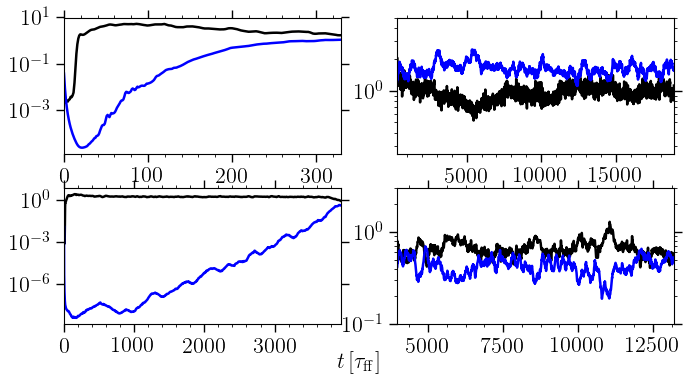}
\caption{Time evolution of the kinetic (black) and magnetic (blue)
  energy densities in the kinematic (left) and saturated (right)
  regimes for simulations A16 with ${\rm Pr_M} = 10$ (top) and A4
  ${\rm Pr_M} = 0.5$ (bottom). {The energies are normalized by
    $E_{\rm kin}.$}}
\label{fig:energies}
\end{figure}

At low ${\rm Pr_M}$ the kinematic, {exponentially growing}, regime
lasts longer than in
the simulations with high ${\rm Pr_M}$. Figure~\ref{fig:energies}
shows the evolution of $E_{\rm mag}$ and $E_{\rm kin}$ in the
kinematic and saturated regimes for runs A4 and A16. The kinematic
regime of simulation A4 lasted {about} 10 times {longer than}
the kinematic regime of run A16. It can be seen that in
simulation A4, $E_{\rm mag}$ is {amplified by six} orders of
magnitude.
In the saturated
regime, both energies {are} comparable {such that} $E_{\rm
  kin}$ is about 1.5 times $E_{\rm mag}$. However, in simulation A16
the kinetic energy density is slightly reduced, while the magnetic
%PJK: \sim => roughly, simplified (but Emag is not bigger than Ekin for all simulations, right?)
energy density is increased by a factor of roughly $1.8$. The $\PrM$
at which $E_{\rm mag}$ overcomes $E_{\rm kin}$ occurs at {\bf ${\rm
    Pr_M} > 1$} for sets A and B, and at {\bf ${\rm Pr_M} > 2$} for
set C. A similar behaviour of the kinetic and magnetic energy
densities was reported before by \cite{browning2008simulations} for
simulations of fully convective stars. In run Cm2 of that work with
${\rm Pr_M} = 5$, $E_{\rm mag}/E_{\rm kin}\leq1$, while Cm with ${\rm
  Pr_M} = 8$ has $E_{\rm mag}/E_{\rm kin}=1.2$.

\begin{figure}[h!]
\centering
\includegraphics[width=\linewidth]{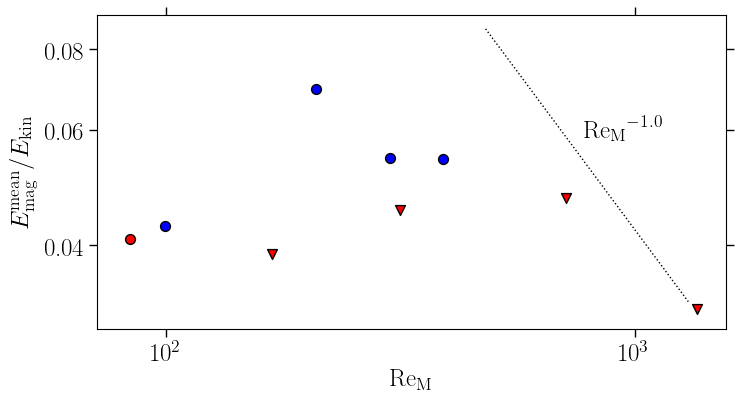}
\caption{Mean magnetic energy normalized by kinetic energy as a function of the magnetic Reynolds number. The
  blue markers are for the last four simulations of set A; the red
  markers are for the simulations of set B. Circles (triangles) are
  for runs with $200^3$ ($576^3$) of resolution. The dotted line corresponds to a power law that shows how a decrease in mean energy with the inverse magnetic Reynolds number might look like.}
\label{fig:mean_emag_rem}
\end{figure}

Table \ref{tab:energies} also includes the energy densities of mean
toroidal ($E_{\rm mag}^{\rm tor}$) and poloidal ($E_{\rm mag}^{\rm
  pol}$) magnetic fields (see columns 8 and 9). $E_{\rm mag}^{\rm
  tor}$ accounts for up to 30$\%$ of total magnetic energy density
and, in general, diminishes as $\ReM$
increases. $E_{\rm mag}^{\rm pol}$ is less than 10$\%$ of $E_{\rm
  mag}$ for almost all simulations. {In general, the 
  %energy of the mean
  ratio of the energy of the mean field to total energy
  %field 
  decreases for high magnetic Reynolds numbers.
  Figure \ref{fig:mean_emag_rem} shows 
  %subsets of simulations from sets
  %A and B %where the mean energy decreases as ${\rm Re_M}$
the saturation level of the mean field as a function of $\ReM$ for subsets of simulations from sets A and B. We do not find a clear trend in the saturation level of the mean energy as a function of the magnetic Reynolds number. 
  %increases for comparable ${\rm Re}$. Two decreasing power-laws were
  %obtained for these models; $\ReM^{-0.86 \pm 0.04}$
  %for set A and $\ReM^{-0.59 \pm 0.06}$ for
  %set B.
  A decrease in the mean energy
  %PJK: But is the decrease catastrophic in our simulations?
  with the inverse
  magnetic Reynolds number is usually associated with catastrophic
  quenching \citep[e.g.][]{CV91,B01}. It can be interpreted as an
  outcome of magnetic helicity conservation, which becomes
  important as $\ReM$ grows
%  \citep[e.g.][]{brandenburg2005astrophysical}. However,
%PJK
%  \cite{hotta2016large} used high resolution simulations in spherical
%  geometry where a small-scale dynamo supressed the small-scale
%  flows. Consequently, the large-scale magnetic field was sustained
%  even at large Reynolds numbers.}
%PJK: On a second thought, I would not cite the Hotta paper at all
%PJK: because of its dubiosity
  \citep[e.g.][]{brandenburg2005astrophysical}.}
{Nevertheless, the boundary conditions in our simulations do allow
  magnetic helicity fluxes}.% still significantly higher magnetic
  %Reynolds numbers may be needed for such fluxes to become effective
  %\citep[e.g.][]{2021PhRvF...6l1701R}.}

Furthermore, the kinetic energy density of the differential rotation,
$E_{\rm kin}^{\rm DR}$, and meridional circulation, $E_{\rm kin}^{\rm
  MC}$, are given in Table \ref{tab:energies}. For simulations with
{a} dynamo $E_{\rm kin}^{\rm DR}$, decreases at higher $\ReM$,
while for simulations {A8 and A12} with no dynamo $E_{\rm
  kin}^{\rm DR}$ is {significantly higher. More specifically the}
runs without a dynamo in set A exhibit roughly five times higher $E_{\rm
  kin}^{\rm DR}$ than runs with a dynamo in the same set. This
{indicates magnetic quenching of differential rotation}. In all of
%PJK
the simulations discussed here, $E_{\rm kin}^{\rm MC}$ is around 1-3
per cent of $E_{\rm kin}$, with the exception of C2, where $E_{\rm
  kin}^{\rm MC} \approx 0.12 E_{\rm kin}$.

\section{Summary and conclusions}\label{conclusions}

We have performed a large sample of simulations of fully convective M
dwarfs using the star-in-a-box setup presented in
\citet{kaepylae2021}. We used the stellar parameters for an M5 dwarf
with 0.21$M_{\odot}$ at three rotation {rates corresponding to
  $\Prot=43$, $61$ and $90$ days},
and varied the magnetic Prandtl number from 0.1 to 10. Our simulations
explore the intermediate to slowly rotating regime. Consistent with
previous work by \citet{kaepylae2021}, we find solar-like differential
rotation in the simulations presented here.

We found different solutions for the large-scale magnetic field
depending on the rotation period and the magnetic Prandtl
number, {which, in our models, fixes the magnetic 
  Reynolds number}. For the simulations with $\tilde{\Omega}=1.0$ {(set A)} and {$\ReM \leq 105$}, the large-scale magnetic
field is cyclic, {with $\Pcyc$ ranging from {309} to
  {471 freefall times}. In this set, we found a slight increase in the
  length of the cycle period with increasing $ \ReM$.} For larger
$\ReM$, no clear cycles are found and instead the behaviour of the
magnetic field and its reversals become irregular. For the simulations
with $\tilde{\Omega}=0.7$ {(set B)} {we found
  cycles for run B1 with $\ReM = 84$, while for higher values of $\ReM$} the
reversals are less {regular}, and instead, a {quasi-static
  configuration} is
found. {For the highest $\ReM$}, the
solutions become irregular. For the case with the lowest rotation rate
($\tilde{\Omega}=0.5$), the field is mainly dipolar for {$\ReM \leq
  750$}. At higher {magnetic Reynolds numbers} the magnetic field
is predominantly irregular and concentrated at mid-latitudes, with
quasi-stationary {fields} near the poles. {We note that in the
  three sets, the large-scale field is irregular and concentrated near
  the equator for the highest ${\rm Pr_M}$ ($\ReM$). Additionally, the
  rms-velocity increases for decreasing rotation for
  comparable $\ReM$.} We also note that for a few
of the simulations, particularly A8, A12 and C2, no dynamo
was found, {because $\ReM$} was below the critical value
to drive a large-scale dynamo.

%{\bf In general, the energy of the mean magnetic field decreases as the magnetic
 % Reynolds number increases, which can interpreted as an outcome of
  %magnetic helicity conservation.} 
  {Furthermore, the ratio
  $\Prot/\Pcyc$ decreases with the Coriolis number, similar to
  the simulations of solar-like stars by
  \cite{strugarek2017reconciling, strugarek2018sensitivity,
    warnecke2018dynamo,
%    viviani2018transition} with simulations of solar-like stars.} Our
%PJK: shorten
    viviani2018transition}.} Our
results confirm the important role of rotation and {dimensionless
  parameters such as $\ReM$ and} the $\PrM$ in determining the
properties of fully convective
dynamos. Depending on the parameters, the magnetic field can show a
clear cyclic behaviour with the cycle period influenced by the
rotation rate and {dimensionless parameters such as {$\ReM\,({\rm Pr_M})$}}. The {large-scale} magnetic field shows
{cycles for low and modest values of $\ReM$ but the cycles are
  lost for highest magnetic Reynolds numbers where irregular or
  quasi-static fields dominate. A similar loss of cyclic solutions was
  reported by \cite{kapyla2017convection} who also increased $\PrM$ to
  increase $\ReM$. Whether the behavior of the dynamo changes if
  $\ReM$ is fixed and $\PrM$ is lowered is yet an open question. This
  is also closer to the parameter regime of late-type stars where
  $\PrM\ll1$ and $\ReM\gg1$, but such parameter regime is extremely
  challenging numerically.}

A very tentative comparison can be pursued with the Proxima Centauri system,
where \citet{klein2021large} inferred a seven-year activity cycle. In
principle the activity cycle inferred in our simulations is in the
range {from five to nine} years, and thus consistent with the
observed data. We note that this comparison is preliminary and even
%though the rotation rate we adopt here is the same, the magnetic
%PJK: is *not* the same, right?
though the rotation rate we adopt here is {similar}, the magnetic
Prandtl number is likely to be different, and even larger differences
concern the magnetic Reynolds number in the star in comparison to the
simulation. As it is well known that the solutions for the magnetic
field depend on these parameters, leading to uncertainty in the
possible interpretation.  It is nevertheless encouraging that the
behaviour found in the simulations is relatively similar.

Overall, the study presented here consists of, to our knowledge, the
largest exploration of the parameter space for dynamo models of fully
convective M dwarfs. Uncertainties remain for instance regarding the
role of the magnetic Reynolds number, which will still be much larger
in realistic systems. While a clear signature of a small-scale dynamo
(SSD) is not found in our simulations, the expectation is that SSD are
present at larger magnetic Reynolds numbers and interact with the
large-scale dynamo, thereby changing the solution.   

\begin{acknowledgements}
CAOR, DRGS and JPH thank for funding via Fondecyt Regular (project
code 1201280). CAOR, DRGS and R.E.M. gratefully acknowledge support by
the ANID BASAL projects ACE210002 and FB210003. DRGS and
R.E.M. gratefully acknowledge support by the FONDECYT Regular
1190621. DRGS thanks for funding via the  Alexander von Humboldt - Foundation, Bonn, Germany. PJK acknowledges finantial support from DFG Heisenberg
programme grant No. KA 4825/4-1. FHN acknowledges financial support by
the DAAD (Deutscher Akademischer Austauschdienst; code 91723643). The
simulations were made using the Kultrun cluster hosted at the
Departamento de Astronomía, Universidad de Concepción, and on HLRN-IV
under project grant hhp00052.

\end{acknowledgements}


\bibliographystyle{aa}
\bibliography{main}

\begin{thebibliography}{48}
\expandafter\ifx\csname natexlab\endcsname\relax\def\natexlab#1{#1}\fi

\bibitem[{{Augustson} {et~al.}(2019){Augustson}, {Brun}, \&
  {Toomre}}]{2019ApJ...876...83A}
{Augustson}, K.~C., {Brun}, A.~S., \& {Toomre}, J. 2019, \apj, 876, 83

\bibitem[{Bice \& Toomre(2020)}]{bice2020probing}
Bice, C.~P. \& Toomre, J. 2020, \apj, 893, 107

\bibitem[{{Brandenburg}(2001)}]{B01}
{Brandenburg}, A. 2001, \apj, 550, 824

\bibitem[{Brandenburg {et~al.}(2017)Brandenburg, Mathur, \&
  Metcalfe}]{brandenburg2017evolution}
Brandenburg, A., Mathur, S., \& Metcalfe, T.~S. 2017, \apj, 845, 79

\bibitem[{Brandenburg {et~al.}(1998)Brandenburg, Saar, \&
  Turpin}]{brandenburg1998time}
Brandenburg, A., Saar, S.~H., \& Turpin, C.~R. 1998, \apj, 498, L51

\bibitem[{Brandenburg \& Subramanian(2005)}]{brandenburg2005astrophysical}
Brandenburg, A. \& Subramanian, K. 2005, \physrep, 417, 1

\bibitem[{Brown {et~al.}(2020)Brown, Oishi, Vasil, Lecoanet, \&
  Burns}]{brown2020single}
Brown, B.~P., Oishi, J.~S., Vasil, G.~M., Lecoanet, D., \& Burns, K.~J. 2020,
  \apjl, 902, L3

\bibitem[{Browning(2008)}]{browning2008simulations}
Browning, M.~K. 2008, \apj, 676, 1262

\bibitem[{{Brun} \& {Browning}(2017)}]{2017LRSP...14....4B}
{Brun}, A.~S. \& {Browning}, M.~K. 2017, Liv. Rev. Sol. Phys., 14, 4

\bibitem[{{Brun} {et~al.}(2004){Brun}, {Miesch}, \& {Toomre}}]{BMT04}
{Brun}, A.~S., {Miesch}, M.~S., \& {Toomre}, J. 2004, \apj, 614, 1073

\bibitem[{{Cattaneo} \& {Vainshtein}(1991)}]{CV91}
{Cattaneo}, F. \& {Vainshtein}, S.~I. 1991, \apjl, 376, L21

\bibitem[{{Chabrier} \& {Baraffe}(1997)}]{1997A&A...327.1039C}
{Chabrier}, G. \& {Baraffe}, I. 1997, \aap, 327, 1039

\bibitem[{{Chandrasekhar}(1961)}]{Ch61}
{Chandrasekhar}, S. 1961, {Hydrodynamic and hydromagnetic stability}

\bibitem[{{Dobler} {et~al.}(2006){Dobler}, {Stix}, \&
  {Brandenburg}}]{dobler2006}
{Dobler}, W., {Stix}, M., \& {Brandenburg}, A. 2006, \apj, 638, 336

\bibitem[{{Featherstone} \& {Hindman}(2016)}]{FH16b}
{Featherstone}, N.~A. \& {Hindman}, B.~W. 2016, \apjl, 830, L15

\bibitem[{{Guerrero} {et~al.}(2019){Guerrero}, {Zaire}, {Smolarkiewicz}, {de
  Gouveia Dal Pino}, {Kosovichev}, \& {Mansour}}]{2019ApJ...880....6G}
{Guerrero}, G., {Zaire}, B., {Smolarkiewicz}, P.~K., {et~al.} 2019, \apj, 880,
  6

\bibitem[{{Hotta} {et~al.}(2022){Hotta}, {Kusano}, \&
  {Shimada}}]{2022ApJ...933..199H}
{Hotta}, H., {Kusano}, K., \& {Shimada}, R. 2022, \apj, 933, 199

\bibitem[{{Jermyn} {et~al.}(2022){Jermyn}, {Anders}, {Lecoanet}, \&
  {Cantiello}}]{2022ApJS..262...19J}
{Jermyn}, A.~S., {Anders}, E.~H., {Lecoanet}, D., \& {Cantiello}, M. 2022,
  \apjs, 262, 19

\bibitem[{{K\"apyl\"a}(2021)}]{kaepylae2021}
{K\"apyl\"a}. 2021, \aap, 651, A66

\bibitem[{K{\"a}pyl{\"a}(2022)}]{kapyla2022solar}
K{\"a}pyl{\"a}, P.~J. 2022, \apjl, 931, L17

\bibitem[{{K{\"a}pyl{\"a}}(2023)}]{2023A&A...669A..98K}
{K{\"a}pyl{\"a}}, P.~J. 2023, \aap, 669, A98

\bibitem[{K{\"a}pyl{\"a} {et~al.}(2020)K{\"a}pyl{\"a}, Gent, Olspert,
  K{\"a}pyl{\"a}, \& Brandenburg}]{kapyla2020sensitivity}
K{\"a}pyl{\"a}, P.~J., Gent, F.~A., Olspert, N., K{\"a}pyl{\"a}, M.~J., \&
  Brandenburg, A. 2020, Geophys. Astrophys. Fluid Dyn., 114, 8

\bibitem[{K{\"a}pyl{\"a} {et~al.}(2017)K{\"a}pyl{\"a}, K{\"a}pyl{\"a}, Olspert,
  Warnecke, \& Brandenburg}]{kapyla2017convection}
K{\"a}pyl{\"a}, P.~J., K{\"a}pyl{\"a}, M., Olspert, N., Warnecke, J., \&
  Brandenburg, A. 2017, \aap, 599, A4

\bibitem[{{K{\"a}pyl{\"a}} {et~al.}(2018){K{\"a}pyl{\"a}}, {K{\"a}pyl{\"a}}, \&
  {Brandenburg}}]{2018AN....339..127K}
{K{\"a}pyl{\"a}}, P.~J., {K{\"a}pyl{\"a}}, M.~J., \& {Brandenburg}, A. 2018,
  Astron. Nachr., 339, 127

\bibitem[{Klein {et~al.}(2021)Klein, Donati, H{\'e}brard, Zaire, Folsom, Morin,
  Delfosse, \& Bonfils}]{klein2021large}
Klein, B., Donati, J.-F., H{\'e}brard, {\'E}.~M., {et~al.} 2021, \mnras, 500,
  1844

\bibitem[{Kochukhov(2021)}]{kochukhov2021magnetic}
Kochukhov, O. 2021, Astron. Astrophys. Rev., 29, 1

\bibitem[{Luukko {et~al.}(2016)Luukko, Helske, \&
  R{\"a}s{\"a}nen}]{luukko2016introducing}
Luukko, P.~J., Helske, J., \& R{\"a}s{\"a}nen, E. 2016, Comput. Stat., 31, 545

\bibitem[{{Morin} {et~al.}(2010){Morin}, {Donati}, {Petit}, {Delfosse},
  {Forveille}, \& {Jardine}}]{2010MNRAS.407.2269M}
{Morin}, J., {Donati}, J.-F., {Petit}, P., {et~al.} 2010, \mnras, 407, 2269

\bibitem[{Moutou {et~al.}(2017)Moutou, H{\'e}brard, Morin, Malo, Fouque,
  Torres-Rivas, Martioli, Delfosse, Artigau, \& Doyon}]{moutou2017spirou}
Moutou, C., H{\'e}brard, E., Morin, J., {et~al.} 2017, \mnras, 472, 4563

\bibitem[{Navarrete {et~al.}(2022)Navarrete, Schleicher, K{\"a}pyl{\"a},
  Ortiz-Rodr{\'\i}guez, \& Banerjee}]{navarrete2022origin}
Navarrete, F.~H., Schleicher, D.~R., K{\"a}pyl{\"a}, P.~J.,
  Ortiz-Rodr{\'\i}guez, C.~A., \& Banerjee, R. 2022, \aap, 667, A164

\bibitem[{Newton {et~al.}(2017)Newton, Irwin, Charbonneau, Berlind, Calkins, \&
  Mink}]{newton2017halpha}
Newton, E.~R., Irwin, J., Charbonneau, D., {et~al.} 2017, \apj, 834, 85

\bibitem[{Parker(1955)}]{parker1955hydromagnetic}
Parker, E.~N. 1955, \apj, 122, 293

\bibitem[{{Pencil Code Collaboration} {et~al.}(2021){Pencil Code
  Collaboration}, {Brandenburg}, {Johansen}, {Bourdin}, {Dobler}, {Lyra},
  {Rheinhardt}, {Bingert}, {Haugen}, {Mee}, {Gent}, {Babkovskaia}, {Yang},
  {Heinemann}, {Dintrans}, {Mitra}, {Candelaresi}, {Warnecke},
  {K{\"a}pyl{\"a}}, {Schreiber}, {Chatterjee}, {K{\"a}pyl{\"a}}, {Li},
  {Kr{\"u}ger}, {Aarnes}, {Sarson}, {Oishi}, {Schober}, {Plasson}, {Sandin},
  {Karchniwy}, {Rodrigues}, {Hubbard}, {Guerrero}, {Snodin}, {Losada},
  {Pekkil{\"a}}, \& {Qian}}]{2021JOSS....6.2807P}
{Pencil Code Collaboration}, {Brandenburg}, A., {Johansen}, A., {et~al.} 2021,
  J. Open Source Softw., 6, 2807

\bibitem[{{Reiners} {et~al.}(2022){Reiners}, {Shulyak}, {K{\"a}pyl{\"a}},
  {Ribas}, {Nagel}, {Zechmeister}, {Caballero}, {Shan}, {Fuhrmeister},
  {Quirrenbach}, {Amado}, {Montes}, {Jeffers}, {Azzaro}, {B{\'e}jar},
  {Chaturvedi}, {Henning}, {K{\"u}rster}, \& {Pall{\'e}}}]{2022A&A...662A..41R}
{Reiners}, A., {Shulyak}, D., {K{\"a}pyl{\"a}}, P.~J., {et~al.} 2022, \aap,
  662, A41

\bibitem[{{Rogachevskii} \& {Kleeorin}(2015)}]{2015JPlPh..81e3904R}
{Rogachevskii}, I. \& {Kleeorin}, N. 2015, , J. Plasma Phys., 81, 395810504

\bibitem[{{Saar} \& {Linsky}(1985)}]{Saar1985}
{Saar}, S.~H. \& {Linsky}, J.~L. 1985, \apjl, 299, L47

\bibitem[{{Schekochihin} {et~al.}(2007){Schekochihin}, {Iskakov}, {Cowley},
  {McWilliams}, {Proctor}, \& {Yousef}}]{SICMPY07}
{Schekochihin}, A.~A., {Iskakov}, A.~B., {Cowley}, S.~C., {et~al.} 2007, New J.
  Phys., 9, 300

\bibitem[{Schrinner {et~al.}(2012)Schrinner, Petitdemange, \&
  Dormy}]{schrinner2012dipole}
Schrinner, M., Petitdemange, L., \& Dormy, E. 2012, \apj, 752, 121

\bibitem[{{Steenbeck} {et~al.}(1966){Steenbeck}, {Krause}, \&
  {R{\"a}dler}}]{1966ZNatA..21..369S}
{Steenbeck}, M., {Krause}, F., \& {R{\"a}dler}, K.-H. 1966, Z. Naturf. A, 21,
  369

\bibitem[{Strugarek {et~al.}(2018)Strugarek, Beaudoin, Charbonneau, \&
  Brun}]{strugarek2018sensitivity}
Strugarek, A., Beaudoin, P., Charbonneau, P., \& Brun, A. 2018, \apj, 863, 35

\bibitem[{Strugarek {et~al.}(2017)Strugarek, Beaudoin, Charbonneau, Brun, \&
  do~Nascimento~Jr}]{strugarek2017reconciling}
Strugarek, A., Beaudoin, P., Charbonneau, P., Brun, A., \& do~Nascimento~Jr,
  J.-D. 2017, Science, 357, 185

\bibitem[{Viviani {et~al.}(2018)Viviani, Warnecke, K{\"a}pyl{\"a},
  K{\"a}pyl{\"a}, Olspert, Cole-Kodikara, Lehtinen, \&
  Brandenburg}]{viviani2018transition}
Viviani, M., Warnecke, J., K{\"a}pyl{\"a}, M.~J., {et~al.} 2018, \aap, 616,
  A160

\bibitem[{Warnecke(2018)}]{warnecke2018dynamo}
Warnecke, J. 2018, \aap, 616, A72

\bibitem[{Wright \& Drake(2016)}]{wright2016solar}
Wright, N.~J. \& Drake, J.~J. 2016, \nat, 535, 526

\bibitem[{Wright {et~al.}(2018)Wright, Newton, Williams, Drake, \&
  Yadav}]{wright2018stellar}
Wright, N.~J., Newton, E.~R., Williams, P.~K., Drake, J.~J., \& Yadav, R.~K.
  2018, \mnras, 479, 2351

\bibitem[{{Yadav} {et~al.}(2015){Yadav}, {Christensen}, {Morin}, {Gastine},
  {Reiners}, {Poppenhaeger}, \& {Wolk}}]{YCMGRPW15}
{Yadav}, R.~K., {Christensen}, U.~R., {Morin}, J., {et~al.} 2015, \apjl, 813,
  L31

\bibitem[{{Yadav} {et~al.}(2016){Yadav}, {Christensen}, {Wolk}, \&
  {Poppenhaeger}}]{2016ApJ...833L..28Y}
{Yadav}, R.~K., {Christensen}, U.~R., {Wolk}, S.~J., \& {Poppenhaeger}, K.
  2016, \apjl, 833, L28

\bibitem[{Yoshimura(1975)}]{yoshimura1975solar}
Yoshimura, H. 1975, \apj, 201, 740

\end{thebibliography}

\end{document}